\documentclass[aps,preprint,nofootinbib,floatfix]{revtex4}
\usepackage{graphicx,color}
\usepackage[latin1]{inputenc}
\usepackage{amsmath,amssymb}
\usepackage{hyperref}
\usepackage{epstopdf}
\usepackage{slashed}
\usepackage{soul}
\usepackage{amssymb}
\usepackage[normalem]{ulem}
\usepackage[caption=false]{subfig}
\usepackage{booktabs}
\usepackage{diagbox}
\usepackage{makecell}
\usepackage{multirow}
\usepackage{enumitem}

\newcommand{\lsim}{\mathrel{\mathop{\kern 0pt \rlap
  {\raise.2ex\hbox{$<$}}}
  \lower.9ex\hbox{\kern-.190em $\sim$}}}
\newcommand{\gsim}{\mathrel{\mathop{\kern 0pt \rlap
  {\raise.2ex\hbox{$>$}}}
  \lower.9ex\hbox{\kern-.190em $\sim$}}}
\newcommand{\sv}{\ensuremath{\langle\sigma v_{\text{rel}}\rangle}}
\newcommand{\svloo}{\ensuremath{\langle\sigma v\rangle_{\texttt{100}}}}

\newcommand{\mev}{\ensuremath{\,\mathrm{MeV}}}
\newcommand{\gev}{\ensuremath{\,\mathrm{GeV}}}

\def  \bcen   {\begin{center}}
\def  \ecen   {\end{center}}
\def  \beq    {\begin{equation}}
\def  \eeq    {\end{equation}}
\def  \beqa   {\begin{eqnarray}}
\def  \eeqa   {\end{eqnarray}}

\def\bea{\begin{eqnarray}}
\def\eea{\end{eqnarray}}

\begin{document}

\title{Exploring sub-GeV dark matter via $s$-wave, $p$-wave, and resonance annihilation with CMB data}
\author{Yu-Ning Wang$^{a,b}$}
\author{Xin-Chen Duan$^{a,b}$}
\author{\\Tian-Peng Tang$^{a}$}
\author{Ziwei Wang$^{a}$}
\email{zwwang@pmo.ac.cn}
\author{Yue-Lin Sming Tsai$^{a,b}$}
\email{smingtsai@pmo.ac.cn}

\affiliation{$^a$Key Laboratory of Dark Matter and Space Astronomy, 
   Purple Mountain Observatory, Chinese Academy of Sciences, Nanjing 210033, China}
\affiliation{$^b$School of Astronomy and Space Science, University of Science and Technology of China, Hefei, Anhui 230026, China}

\date{\today}

\begin{abstract}
We revisit constraints on sub-GeV dark matter (DM) annihilation via $s$-wave, $p$-wave, and resonance processes using current and future CMB data from Planck, FIRAS, and upcoming experiments such as LiteBIRD, CMB-S4, PRISTINE, and PIXIE.
For $s$-wave annihilation, we provide updated limits for both $e^{+}e^{-}$ and $\pi\pi$ channels, with the profile likelihood method yielding stronger constraints than the marginal posterior method. 
In the $p$-wave case, we comprehensively present a model-independent inequality for the 95\% upper limits from FIRAS, PRISTINE, and PIXIE, with future experiments expected to surpass current BBN limits. 
For resonance annihilation, 
we report---for the first time---the $95\%$ upper limits on the decay branching ratio of the mediator particle, 
when the resonance peaks during the recombination epoch and at higher redshifts. 
Overall, our study highlights the complementary strengths of spectral distortions and CMB anisotropies in probing sub-GeV DM annihilation.

\end{abstract}

\maketitle
\section{Introduction \label{sec:intro}}

Dark matter (DM) is widely acknowledged as a fundamental constituent of the Universe and a primary agent in the formation of cosmological structures. However, its nature remains an elusive mystery.
The observed DM relic density provides an important window into understanding the nature of DM.
Among the various mechanisms proposed to account for the relic density, the weakly interacting massive particle as thermally produced via the freeze-out mechanism from the Standard Model (SM) plasma in the early Universe is the most compelling scenario~\cite{Goldberg:1983nd, Goodman:1984dc, Blumenthal:1984bp} (see~\cite{Jungman:1995df, Bergstrom:2000pn, Bertone:2004pz} as reviews).
Canonical models, such as supersymmetry, predict weakly interacting massive particles with masses around $\mathcal O(100)$ GeV~\cite{Ellis:1983ew}, but decades of direct detection experiments, including XENON~\cite{Aprile:2012zx, PhysRevLett.131.041003}, PandaX~\cite{PandaX-II:2017hlx, PhysRevLett.127.261802}, DarkSide-50~\cite{DarkSide-50:2023fcw}, LZ~\cite{PhysRevLett.131.041002} and CDEX~\cite{CDEX:2018lau, PhysRevLett.129.221301}, have yielded no conclusive signals, compressing the parameter space for massive DM models~\cite{Akerib2022Snowmass2021CF}.

The absence of discoveries has spurred critical reassessments of both experimental methodologies and theoretical assumptions on light DM consisting of particles in the MeV-GeV range.
For light DM, a universal bound of light DM mass was originally studied by Lee and Weinberg~\cite{Lee:1977ua} and concluded that if the mediator of the DM-SM interaction is part of the SM particles, the DM mass would have to be greater than a lower bound of the order of $2\gev$. 
To beat the bound, except for non-thermal freeze-out mechanisms~\cite{Hall:2009bx, Bernal:2017kxu, Dvorkin:2020xga} and asymmetric DM models~\cite{Kaplan:2009ag, Zurek:2013wia}, another typical way to escape the Lee-Weinberg constraints is to study light DM by introducing new ``dark sector'' particles, i.e., new force mediators and matter fields other than DM.
For a detailed review, we refer readers to Ref.~\cite{Lin:2019uvt}. 
To search for sub-GeV DM particles with masses above the MeV scale, indirect detection experiments utilize the cosmic ray detector Voyager-1~\cite{doi:10.1126/science.1236408} and various gamma-ray telescopes, including Fermi-LAT~\cite{Ajello_2021}, COMPTEL~\cite{article}, INTEGRAL-SPI~\cite{Siegert2015GammaraySO}, VLAST~\cite{Pan:2024adp}, GECCO~\cite{Moiseev:2023zkv}, and GRAMS~\cite{Aramaki:2023nrm} (see~\cite{Bertuzzo:2017lwt, ODonnell:2024aaw} for reviews) to probe annihilation or decay signals in the Galactic Center and dwarf galaxies~\cite{DelaTorreLuque_2024, DelaTorreLuque:2024wfz, PhysRevD.103.063022, Aghaie:2025dgl}.

The Cosmic Microwave Background (CMB) serves as a critical probe for MeV-GeV DM due to its sensitivity to energy injection during recombination~\cite{PhysRevD.88.063502, PhysRevD.95.023010, PhysRevD.76.061301, PhysRevD.70.043502, PhysRevD.80.023505, Slatyer:2009yq, Poulin_2017, PhysRevD.72.023508, 10.1143/PTP.123.853, PhysRevD.84.027302,Diamanti2013ConstrainingDM} and alterations in the relativistic energy density $N_{\rm eff}$~\cite{Sabti:2019mhn, Lin:2011gj}. 
In the thermal history of the universe, DM annihilation generates high-energy photons and electrons that heat and ionize hydrogen and helium gas, broadening the last scattering surface.
This results in a relative suppression of temperature fluctuations accompanied by an enhancement of polarization in the CMB~\cite{PhysRevD.72.023508}.
Meanwhile, deviations of the CMB frequency spectrum, known as blackbody spectral distortions of CMB~\cite{Zeldovich:1969ff, Sunyaev:1970bma, Burigana:1991eub, Hu:1992dc, Chluba:2011hw, Chluba:2012gq}, provide a complementary probe to study energy injection processes in the early universe.
These effects can be quantitatively constrained through bounds on the thermally-averaged $s$-wave annihilation cross-section $\sv$ into SM final states.
For DM annihilating into electromagnetically charged particles, observations from the Planck satellite during the recombination epoch establish a limit of $\sv_{\mathrm{CMB}} \lesssim 3\times 10^{-26} \mathrm{cm}^3 \mathrm{s}^{-1} (m_\chi /10\gev) $~\cite{Planck:2018vyg}, where $m_\chi$ is DM mass.
Given that standard thermal relic DM scenarios predict a thermally-averaged annihilation cross-section $\sv \sim 10^{-26} \mathrm{cm}^3 \mathrm{s}^{-1}$ during the freeze-out epoch, pure $s$-wave annihilation scenario is excluded for DM masses $m_\chi \lesssim 10 \gev$.
However, when accounting for velocity-dependent annihilation cross-sections, 
the approximation $\sv \sim 10^{-26} \, \mathrm{cm}^3 \, \mathrm{s}^{-1}$ no longer holds.
Also, $\sv$ obtained from precise relic density calculations~\cite{Chatterjee:2025vdz,Duan:2024urq,Binder:2017rgn} cannot be directly related to $\sv_{\mathrm{CMB}}$.
Furthermore, by incorporating velocity dependence, the annihilation rate during the recombination epoch can be suppressed due to the reduced relative velocities of DM particles, thereby evading constraints from CMB observations.
Among these mechanisms, $p$-wave annihilation (velocity-squared dependent)~\cite{An:2016kie, BOEHM2004219,Kumar:2013iva} and resonant enhancement scenarios~\cite{PhysRevD.79.095009,PhysRevD.43.3191} have emerged as viable solutions in recent studies.

In this article, we investigate the constraints on sub-GeV DM annihilation imposed by the data of COBE/FIRAS~\cite{Fixsen:1996nj, Mather:1998gm}, Planck satellite~\cite{Planck:2018vyg, Planck:2019nip} and Baryon Acoustic Oscillations (BAO)~\cite{10.1093/mnras/stx721, Buen-Abad_2018, 10.1111/j.1365-2966.2011.19250.x, 10.1093/mnras/stv154}, considering both $e^{+}e^{-}$ and $\pi\pi$ final states through $s$-wave, $p$-wave, and resonant processes.
We utilize the \texttt{Hazma} tool~\cite{Coogan:2019qpu} to generate precise photon and electron spectra, and calculate the energy deposition into the intergalactic medium (IGM) using \texttt{DarkHistory v2.0}~\cite{Liu:2019bbm}.
This systematic approach allows us to probe the nature of light DM and its interactions with greater precision, offering new insights into the phenomenology of sub-GeV DM annihilation.

This paper is organized as follows. In Section~\ref{sec:DM Model} we describe three DM annihilation scenarios while the energy deposition into the background in Section~\ref{sec:DM energy injection}.
We outline the CMB constraints of blackbody spectral distortions in Section~\ref{sec: BSD} and CMB anisotropies in Section~\ref{sec:powerspectra}. Additionally, in Section~\ref{sec:statistics}, we present the statistical framework used in our analysis.
We present the results in Section~\ref{sec:Result} and discussion in Section~\ref{sec:summary}. In this paper, we use natural units $c=\hbar=k_\text{B}=1$.

\section{The DM annihilation induced energy deposition}
\label{sec:DMann}

DM annihilation can inject energy into CMB photons and gas, thereby affecting measurements of CMB anisotropies. 
The various velocity dependence and annihilation final states lead to different energy injection profiles. 
In this section, we begin by examining the velocity dependence of the annihilation cross-section and then detail our calculations of the effects of DM energy deposition on the evolution of ionization fraction and gas temperature.

\subsection{DM annihilation scenarios}
\label{sec:DM Model}

In addition to annihilation processes involving resonance and Sommerfeld enhancements, the DM annihilation amplitude squared $|\mathcal{M}|^2$ can be expanded in terms of the DM velocity $v_{\chi}$ as $|\mathcal{M}|^2 = a_s + b_p v_{\chi}^2 + \mathcal{O}(v_\chi^4)$, where the terms $a_s$ and $b_p v_{\chi}^2$ correspond to $s$-wave and $p$-wave contributions, respectively. 
In the case of resonance annihilation mediated by a particle $\phi$ of mass $m_\phi$, its $|\mathcal{M}|^2$ is inversely proportional to the propagator, 
$\{(E_\text{cm}^2 - m_\phi^2)^2 + (m_\phi \Gamma_\phi)^2\}$. 
When DM mass $m_\chi\approx m_\phi/2$ and the decay width $\Gamma_\phi\ll m_\phi$, the annihilation cross-section can be boosted by $v_{\rm rel}^{-4}$. 
We detail each type of annihilation based on its distinct velocity dependence.
\begin{itemize}
    \item \textbf{$s$-wave annihilation}:\\
In this scenario, $|\mathcal{M}|^2$ is velocity-independent, leading to a constant velocity-averaged cross-section $\sv \propto a_s$.
Thus, $\sv$ is also independent of redshift $z$, and DM mass $m_\chi$ and $\sv$ are free parameters.

    \item \textbf{$p$-wave annihilation}:\\
In this scenario, $|\mathcal{M}|^2$ is velocity-dependent.
Consequently, the DM kinetic decoupling temperature from the SM plasma, $T_\text{kd}$, becomes a crucial input in shaping the evolution of DM velocity, which in turn impacts $\sv$.
In Appendix~\ref{app:p-wave}, we derive the root-mean-square velocity of DM particles as   
\begin{equation}\label{equ:p-wave}
\begin{aligned}
    v_{\text{rms}}^2\equiv \langle v_\chi^2 \rangle = 
    1.66\times10^{-19} \times  (1+z)^2
    \left(\frac{g_{*,\text{kd}}}{g_{*,0}}\right)^{-2/3}
    \left(\frac{1~\text{MeV}}{m_\chi}\right)
    \left(\frac{1~\text{MeV}}{T_{\text{kd}}}\right),
\end{aligned}
\end{equation}
where $g_{*,\text{kd}}$ and $g_{*,0}$ are the effective numbers of relativistic degrees of freedom at the redshift of DM kinetic decoupling and at present $(z=0)$, respectively. For $T_{\rm kd}<0.1~\text{MeV}$, the ratio $g_{*,\text{kd}}/g_{*,0}$ is equal to 1, whereas for $T_\text{kd}\sim \mathcal{O}(10~\text{MeV})$ to $T_\text{kd}\sim \mathcal{O}(1~\text{MeV})$, it is approximately 3.18.
The quantity $\langle v_\chi^2 \rangle$ depends on the DM temperature $T_\chi$, which is a function of redshift $z$, $m_\chi$, and $T_{\rm kd}$. 

Finally, by using Eq.~\eqref{equ:p-wave}, we define a free parameter $b$ with the same units of $\sv$ in this paper $\sv \equiv b \langle v_\chi^2 \rangle$,
thus there are three free parameters in the dark sector: \{$m_\chi$, $b$, and $T_{\rm kd}$\}.

\begin{figure*}[ht]
\centering
\includegraphics[width=8cm]{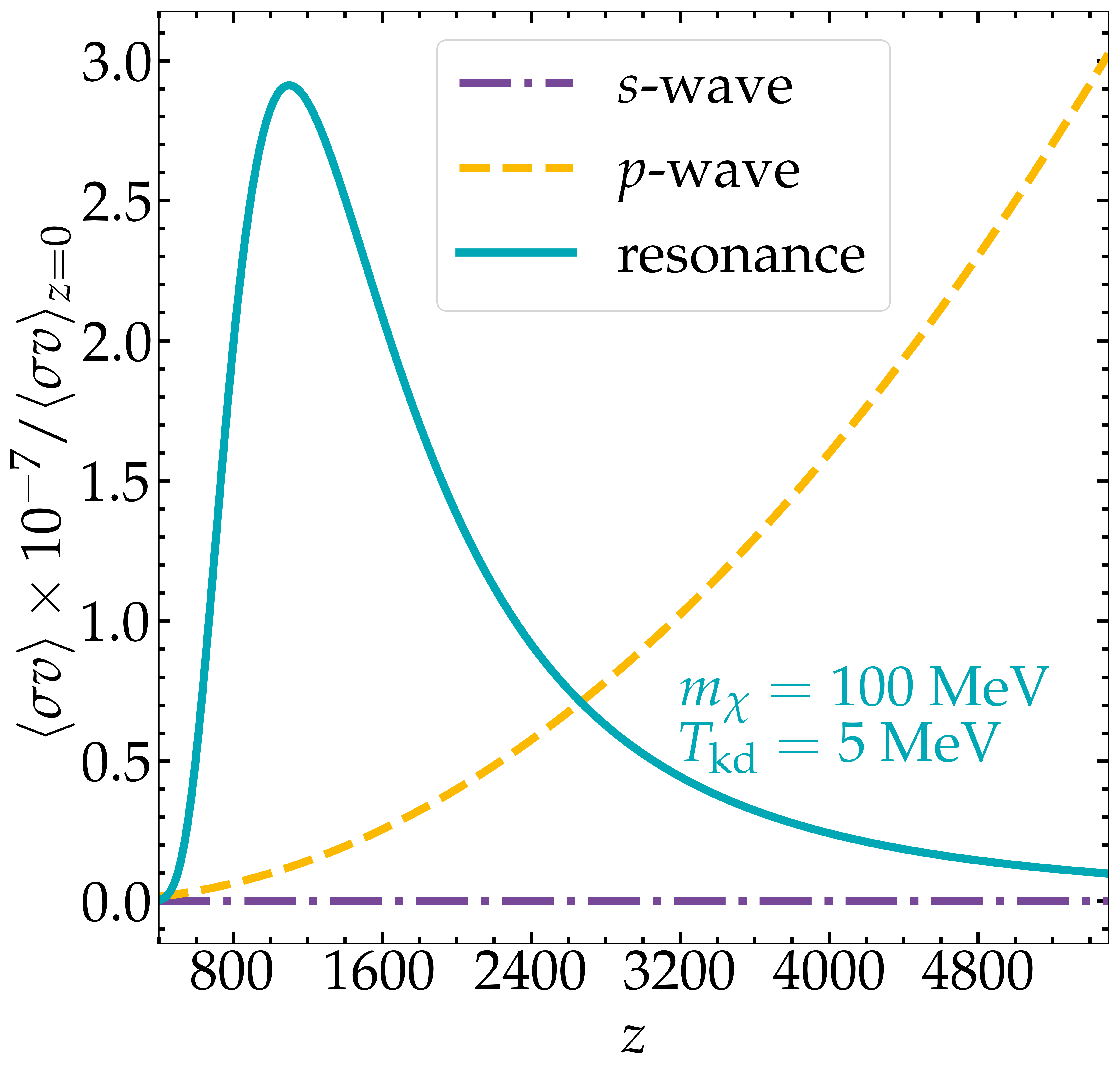}
\includegraphics[width=8cm]{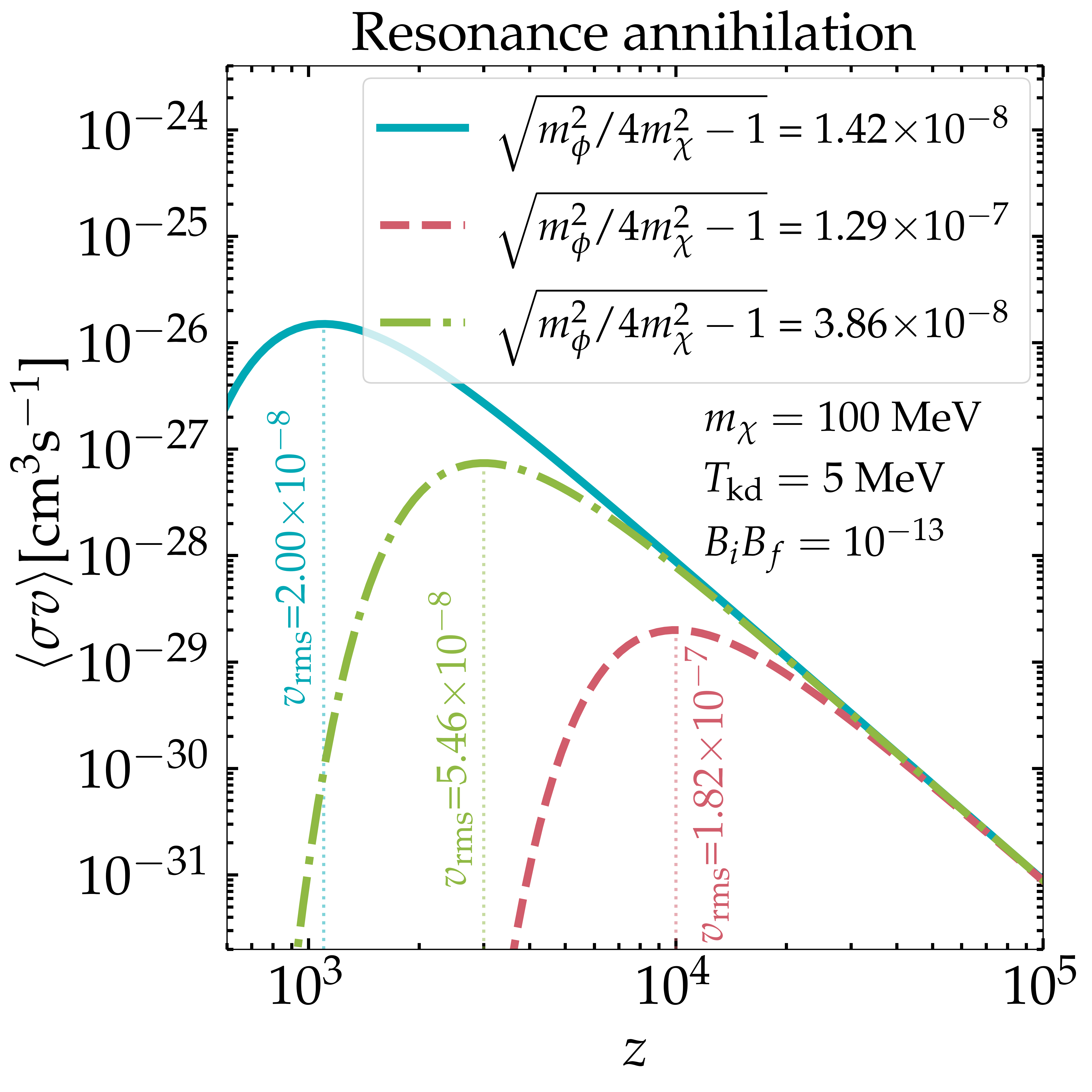}
 \caption{
 The evolution of annihilation cross-section for a comparison between three scenarios (left panel) and three different $z_{\rm peak}$ benchmarks in the resonance scenario (right panel).   
 Left panel: Thermal averaged cross-section $\sv$ normalized to the cross-section at $z=0$, with respect to redshift. 
 The free parameters associated with $s$-wave and $p$-wave interactions are eliminated, 
 while the resonance scenario is characterized by $m_\chi = 100~\text{MeV}$ and $T_{\text{kd}} = 5~\text{MeV}$.
 Right panel: The dashed, dash-dotted,  and solid lines represent $z_{\text{peak}} = \{10^4, 3000, 1100\}$, respectively. }
 \label{fig:cross-section}
\end{figure*}

    \item \textbf{Resonance annihilation}:\\
For an annihilation $\chi \chi\to \phi\to f\bar f$ where the SM fermion is denoted as $f$, 
the cross-section with resonance condition $2 m_\chi\approx m_\phi$ can be well described by   
the Breit-Wigner formula~\cite{PhysRevD.79.095009},  
\begin{equation}\label{equ:resonance sigma}
    \sigma=\frac{16\pi}{E^2_{\text{cm}}\Bar{\beta_i}\beta_i}\frac{m_\phi^2\Gamma_\phi^2}{(E^2_{\text{cm}}-m_\phi^2)^2+m_\phi^2\Gamma_\phi^2}B_i B_f, 
\end{equation}
where $E_\text{cm}$ is the center-of-mass energy, while $\Gamma_\phi$ is the decay width of the mediator $\phi$. 
To escape from the constraints from Big Bang Nucleosynthesis (BBN), we set $\Gamma_\phi=10^{-21}~\text{MeV}$, corresponding to the life time $\tau_\phi\sim 1$~sec. 
The initial state phase space factors are given by $\Bar{\beta_i} = \sqrt{1 - 4m_\chi^2/m_\phi^2}$ and $\beta_i = \sqrt{1 - 4m_\chi^2/E_{\text{cm}}^2}$, while the decay branching ratios for $\phi \to \chi\chi$ and $\phi \to f\bar{f}$ are denoted as $B_i$ and $B_f$, respectively.
For convenience, we introduce a parameter $\xi$ to describe the resonance, defined by
\begin{equation}\label{equ:resonance condition}
    m_\phi^2=4m_\chi^2(1+\xi),~{\rm with }~\xi\ll 1.
\end{equation}

We can see that $\xi>0$ implies $m_\phi>2m_\chi$. 
Using Eq.~\eqref{equ:resonance condition} and $\gamma\equiv\Gamma_\phi/m_\phi$, we can further rewrite Eq.~\eqref{equ:resonance sigma} as
\begin{equation}
    \sigma=\frac{16\pi}{m_\phi^2\Bar{\beta_i}\beta_i}\frac{\gamma^2}{(-\xi+v_{\text{rel}}^2/4)^2+\gamma^2}B_i B_f.
\end{equation}
Considering that DM are non-relativistic at the resonance, we can adopt the Maxwell-Boltzmann velocity distribution for DM to compute the velocity-averaged annihilation cross-section
\begin{equation}\label{equ:resonance thermal-averaged}
    \sv=\frac{1}{(2\pi v^2_{\text{rms}}/3)^3}\int \text{d}\Vec{v}_1\int \text{d}\Vec{v}_2~e^{-3(v_1^2+v_2^2)/(2v^2_{\text{rms}})}\times \sigma \times \left|\Vec{v}_1-\vec{v}_2\right|,
\end{equation}
where the root-mean-square velocity $v_{\text{rms}}^2=\langle v^2_\chi\rangle=3T_\chi/m_\chi$ derived in \eqref{equ:vrms}.
Since the energy injection from the non-resonant component is negligible, we focus on the velocity-averaged cross-section near resonance. 
By applying the narrow width approximation, we derive the velocity-averaged cross-section near resonance (see App.~\ref{app:resonance} for details),
\begin{equation}\label{equ:thermal resonance}
    \sv=
    \frac{576}{\sqrt{3}}
        \frac{\pi^{3/2}}{m_\phi^2}
        \frac{\gamma}{v^3_{\text{rms}}}
        \exp{[-3\xi/v^2_{\text{rms}}}]
        B_i B_f. 
\end{equation}

Interestingly, together with Eq.~\eqref{equ:p-wave}, we notice the peak of Eq.~\eqref{equ:thermal resonance} located at
\begin{equation}\label{equ:z peak}
    z_\text{peak}=\frac{\sqrt{2\xi}}{4.07\times 10^{-10}}\sqrt{
    \Big(\frac{m_\chi}{1~\text{MeV}}\Big)\Big(\frac{T_{\text{kd}}}{1~\text{MeV}}\Big)}.
\end{equation}

Therefore, once $z_{\rm peak}$ is fixed, the remaining free parameters for the $\phi$ resonance scenario are \{$m_\chi$, $T_{\rm kd}$, and $B_i B_f$\}, 
where we treat $B_i B_f$ as a single combined parameter in this work.

\end{itemize}

In Fig.~\ref{fig:cross-section}, we compare the evolution of the annihilation cross-section for three scenarios in the left panel and for three different $z_{\rm peak}$ benchmarks of the resonance scenario in the right panel. 
In the left panel, we normalize $\sv$ by its value at $z = 0$, 
thus the $s$-wave scenario (purple dash-dotted line) indicates $\sv/\sv_{z=0} = 1$, 
while $\sv/\sv_{z=0} = (1+z)^2$ is obtained by the $p$-wave scenario (yellow dashed line). 
Intuitively, the $p$-wave scenario is expected to be easier to detect at larger $z$. 
The cyan solid line represents the resonance scenario with $z_{\rm peak} = 1100$. 
Clearly, if $z_{\rm peak}$ occurs during the recombination epoch, the resonance scenario offers a more promising detection prospect than the other two.

In the right panel, we illustrate the resonance scenario with $m_\chi = 100 \, \mathrm{MeV}$, $T_{\rm kd}=5\mev$, and $B_i B_f=10^{-13}$ as an example. 
We present three benchmarks: $z_{\text{peak}} = 10^4$ (red dashed line), $z_{\text{peak}} = 3000$ (green dash-dotted line),
and $z_{\text{peak}} = 1100$ (cyan solid line), corresponding to $v_{\rm rms} = 1.82 \times 10^{-7}$, $v_{\rm rms} = 5.46 \times 10^{-8}$, and $v_{\rm rms} = 2.0 \times 10^{-8}$, respectively. 

We find that achieving resonance in the range $600 < z_{\rm peak} < 2\times 10^6$ requires fine-tuning the DM velocity to $v_\text{rms} \lesssim \mathcal{O}(10^{-4})$. This poses a challenge for indirect detection, as the DM dispersion velocity in halos $v_\text{rms} \sim \mathcal{O}(10^{-3})$.

\subsection{Energy Deposition}
\label{sec:DM energy injection}

When DM particles annihilate into primary particles, such as SM particles and long-lived new particles, 
these can decay into stable particles and inject energy into the universe, 
while neutrino final states do not contribute to gas heating or ionization. 
For DM masses from a few MeV to a few GeV, possible annihilation channels include two-body final states (e.g., $2e^\pm$, $2\mu^\pm$, or $2\pi$) or four-body final states (e.g., $4e^\pm$, $4\mu^\pm$, or $4\pi$), which arise from the decay of two long-lived new particles. 
However, Planck constraints on four-body models are similar to those on two-body models across a wide DM mass range, 
as the constraints are mainly sensitive to the total energy of $e^\pm$ and $\gamma$ final states, not their spectrum shapes~\cite{Clark2017DarkMA}. 
In this work, we focus on two-body final states (electron-positron pairs and pions), since electrons and positrons are leptons, while pions are mesons made of up and down quarks. 
For simplicity, we adopt the Higgs portal model to compute the pion branch ratios, $\text{BR}(\chi\chi\rightarrow \pi^+\pi^-) = 2/3$ and $\text{BR}(\chi\chi\rightarrow \pi^0\pi^0) = 1/3$.

We compute DM energy deposition that goes into ionization, heating and excitation of the IGM, as 
\begin{equation}
    \bigg(\frac{\text{d}E}{\text{d}V~\text{d}t}\bigg)^{\text{dep}}_c=f_c(z,\mathbf{x})\bigg(\frac{\text{d}E}{\text{d}V~\text{d}t}\bigg)^{\text{inj}},
\end{equation}
where the energy injections induced by DM annihilation are 
\begin{equation}
    \bigg(\frac{\text{d}E}{\text{d}V~\text{d}t}\bigg)^{\text{inj}}=\Omega_\chi^2\rho_{\rm crit}^2(1+z)^6\langle\sigma v_{\text{rel}}\rangle/m_\chi,
\end{equation}
where $V$ is comoving volume, $t$ is physical time, $\Omega_\chi$ is DM density parameter of today, $\rho_{\rm crit}$ is the critical density of today. 
The index $c = \{\text{H ion, heat, Ly}\alpha\}$ denotes hydrogen ionization, heating, or Ly$\alpha$ excitation channels. 
The deposition function $f_c(z, \mathbf{x})$, dependent on redshift $z$ and ionization fractions $\mathbf{x}\equiv \{x_\text{HII}, x_\text{HeII}, x_\text{HeIII}\}$, 
is computed via \texttt{DarkHistory}~\cite{Liu:2023nct,Liu:2023fgu} using photon and electron spectra from DM annihilation generated by \texttt{Hazma}~\cite{Coogan_2020}.

The impact of energy injection from DM annihilation on the ionization history is given in Ref.~\cite{Liu:2019bbm} by the following equations:
\begin{equation}
\begin{aligned}\label{dTdx}
    \dot T^{\text{inj}}_m &= \frac{2 
    f_{\text{heat}} (z,\textbf{x})
    }{3(1+\mathcal{F}_\text{He}+x_e)n_\text{H}}\left(\frac{\text{d}E}{\text{d}V~\text{d}t}\right)^{\text{inj}}, \\
    \dot x^{\text{inj}}_\text{HII} &= 
    \frac{f_{\text{H ion}}(z,\mathbf{x})}{\mathcal{R}n_\text{H}} \left(\frac{\text{d}E}{\text{d}V~\text{d}t}\right)^{\text{inj}},
\end{aligned}
\end{equation}
where $\mathcal{F}_\text{He} \equiv n_\text{He}/n_\text{H}$ is the helium-to-hydrogen ratio, and $n_\text{He}$ and $n_\text{H}$ are the number densities of helium and hydrogen (both neutral and ionized), respectively. The ionization potential of hydrogen is $\mathcal{R} = 13.6~\text{eV}$, and $\mathcal{C}$ is the Peebles C factor, which represents the probability of a hydrogen atom in the $n=2$ state decaying to the ground state before photoionization occurs~\cite{PhysRevD.83.043513,osti_4507738}.

We use $\texttt{DarkHistory v2.0}$~\cite{Liu:2023nct,Liu:2023fgu}, 
which improves the treatment of low-energy photons and electrons compared to $\texttt{DarkHistory v1.0}$~\cite{Liu:2019bbm}, 
to compute additional energy injection from DM annihilation into the universe thermal and ionization evolution.
After tabulating the results of Eq.~\eqref{dTdx}, we insert them into $\texttt{CLASS}$~\cite{Lesgourgues:2011re,DiegoBlas_2011,Lesgourgues:2011rg,Lesgourgues:2011rh} to calculate the CMB power spectra and blackbody spectral distortions.
We list the configuration parameters for \texttt{DarkHistory} and \texttt{CLASS} in Appendix~\ref{app:code}.

\section{Constraints from CMB}
\label{sec:Constraints}

This section aims to briefly outline the CMB constraints and the statistical framework used to explore the parameter space of three benchmark scenarios.

\subsection{Blackbody spectral distortions}
\label{sec: BSD}

In the early universe, photons are thermalized efficiently because of rapid interactions with baryons, including double Compton scattering ($e^- + \gamma \rightarrow e^- + \gamma +\gamma$), bremsstrahlung ($e^- + N \rightarrow e^- + N + \gamma$), and Compton scattering ($\gamma + e^- \rightarrow \gamma + e^-$), which leads to blackbody spectral distortions
\begin{equation}
    I(E_{\gamma},T_z) = \frac{E_{\gamma}^3}{\pi^3}\frac{1}{e^{E_\gamma/T_z} -1} \propto \frac{1}{e^{x}-1},\,
    \label{eq:BB_spe}
\end{equation}
where $x\equiv E_{\gamma}/T_z$ and  $T_z\equiv T_{\text{CMB,0}}(1+z)$ represent the reference temperature in equilibrium with the thermal bath.\footnote{Radiation actual temperature $T_\gamma$ may differ from $T_z$ when energy transfer happens during photon scattering with electrons. However, it is just the temperature deviation from the reference temperature, which is eliminated by coinciding the reference temperature at $z=0$ with the observation.} 
In an expanding universe, above interactions weaken over time and become insufficient before recombination, 
allowing DM-induced energy injections to leave imprints on the CMB frequency spectrum~\cite{10.1093/mnras/stt1733, Lucca:2019rxf, Lucca:2023seh, Li2024ObservabilityOC}.

As $z\lsim z_\text{th} \approx 2\times 10^{6}$, the injected energy cannot fully thermalize, and the processes changing the number density such as 
double Compton scattering and bremsstrahlung become inefficient. The chemical potential of photons becomes non-zero, proportional to the energy deposit, which creates $\mu$-distortion. By perturbing the blackbody spectrum Eq.~\eqref{eq:BB_spe}, the $\mu$-distortion of CMB is defined as follows,
\begin{equation}
    I(E_{\gamma},x;\mu) =   \frac{E_{\gamma}^3}{\pi^3}\frac{1}{e^{x+\mu}-1}\simeq \frac{E_{\gamma}^3}{\pi^3}\left(\frac{1}{e^x-1}-\mu\frac{e^x}{(e^x-1)^2}\right).\,
    \label{eq:mu_defin}
\end{equation}

The $\mu$-distortion from energy injection of DM annihilation is evaluated from Green's function method~\cite{Chluba:2013vsa,Chluba:2015hma}
\begin{equation}
    \mu = 1.401 \int_{0}^{\infty}
    \frac{f_\text{heat}(z)}{\rho_\gamma(z)}\bigg(\frac{\text{d}E}{\text{d}V~\text{d}t}\bigg)^{\text{inj}}\frac{1}{(1+z)H(z)}\cdot \mathcal J_\mu(z) \text{d}z,
\end{equation}
where $\rho_\gamma\approx 0.26(1+z)^4~\text{eV/cm}^3$. 
The Green's function is evaluated as  
\begin{equation}
\mathcal J_{\mu}(z) \approx 
\left\{1 - \exp{\left[-\left(\frac{1+z}{5.8\times10^4}\right)^{1.88}\right]}\right\}
\exp{\left[-\left(\frac{z}{z_{\text{th}}}\right)^{5/2}\right]}.
\end{equation}
Since the universe is fully ionized, and the energy injection follows the on-the-spot approximation~\cite{PhysRevD.72.023508,Slatyer:2009yq}. 
Under such an assumption, DM annihilation into photons and electrons results in instantaneous and complete energy deposition into heating the IGM.

When $z\lsim z_{\mu y}\approx 5 \times 10^4$, Compton scattering as well as energy redistribution become inefficient, driving photons away from Bose-Einstein distribution, which creates $y$-distortion. 
By solving Kompaneets equation \cite{1957JETP....4..730K} and expanding the distribution function with small deviation from blackbody spectrum, the $y$-distortion of CMB is defined as,
\begin{equation}
    I(E_{\gamma},x;y) \simeq \frac{E_{\gamma^3}}{\pi^3}\frac{1}{e^x -1}\left(1 + y \frac{xe^x}{e^x-1}\left(x\coth\left(\frac{x}{2}\right)-4\right)\right).
\end{equation}
To estimate the $y$-distortion from early energy release, we exclude contributions after recombination $z_\text{rec}\approx 1000$, using the Green's function method~\cite{Chluba:2013vsa,Chluba:2015hma}
\begin{equation}
    y = \frac{1}{4} \int_{z_\text{rec}}^{\infty}
    \frac{f_\text{heat}(z)}{\rho_\gamma(z)}\bigg(\frac{\text{d}E}{\text{d}V~\text{d}t}\bigg)^{\text{inj}}\frac{1}{(1+z)H(z)}\cdot \mathcal J_y(z) \text{d}z,
\end{equation}
where 
\begin{equation}
    \mathcal J_{y}(z) \approx
    \left( 1+ \left[\frac{1+z}{6\times 10^4}\right]^{2.58} \right)^{-1},
    ~ z \geq z_\text{rec}\simeq 10^3 .
\end{equation}

The FIRAS experiment aboard the COBE satellite provided one of the most precise measurements of the CMB spectrum. 
It reported $|\mu| < 4.7 \times 10^{-5}$~\cite{PhysRevD.106.063527}, 
improving the previous monopole $\mu$-distortion limit $|\mu| < 9 \times 10^{-5}$~\cite{Fixsen_1996}, 
due to more robust foreground cleaning and $|y|<1.5\times 10^{-5}$~\cite{Fixsen_1996}.

Future missions aim to refine these limits and potentially detect smaller distortions. 
The 95\% upper limit for $\mu$ from PRISTINE will be $|\mu| < 8 \times 10^{-7}$~\cite{Chluba2019NewHI}, 
while PIXIE will set a more stringent limit of $|\mu| < 8 \times 10^{-8}$~\cite{Chluba2019NewHI} at the same confidence level.\footnote{Super-PIXIE allows for potential detection of $\mu$-distortion values as low as $|\mu| \approx 2 \times 10^{-8}$ at $3\sigma$ confidence level~\cite{Chluba2019NewHI,10.1093/mnras/stw945}; the constraint from Super-PIXIE is not considered in this study.}
In this work, we do not include projected $y$-distortion limits for the following reasons. 
While future experiments like PIXIE are expected to achieve exceptional sensitivity to Compton $y$-distortions, 
with projected limits of $y < 2 \times 10^{-9}$~\cite{A.Kogut_2011}, 
these limits are still much lower than the predicted reionization signal. 
Including reionization would require accounting for theoretical uncertainties from Compton scattering off electrons heated by UV radiation, supernovae, and shocks, which are estimated to be around $\mathcal{O}(10^{-6})$~\cite{10.1093/mnras/stw945}. 
Future improvements in constraints from high-sensitivity experiments depend critically on a detailed understanding of reionization. 

The current $y$-distortion constraint ($|y| < 1.5\times 10^{-5}$) is an order of magnitude larger than the anticipated $y$-distortion from reionization. Therefore, in this work, we concentrate on contributions from $z > 1000$, excluding the reionization contribution ($z \lesssim 10$), and refrain from using projected $y$-distortion limits to probe DM annihilation.

\subsection{CMB anisotropies}
\label{sec:powerspectra}

DM annihilation produces high-energy photons and electrons that heat and ionize hydrogen gas, broadening the last scattering surface. 
This increases the residual free electron fraction after recombination, enhancing Thomson scattering. 
As a result, the damping tail of the CMB temperature power spectrum at small angular scales is suppressed due to photon redistribution~\cite{PhysRevD.72.023508,PhysRevD.80.023505,Slatyer:2009yq}. 
Furthermore, increased ionization at later times boosts the low-$\ell$ polarization power spectrum 
by enhancing the visibility of reionization-induced polarization~\cite{PhysRevD.72.023508}. 

As shown in Ref.~\cite{Huang:2021voi}, the current Planck 2018 angular power spectra data~\cite{Planck:2018vyg} is particularly sensitive to DM-induced energy injection in the redshift range of 600 to 1000. 
This Planck data includes the baseline high-$\ell$ power spectra (\texttt{TT}, \texttt{TE}, and \texttt{EE}), the low-$\ell$ power spectrum \texttt{TT}, the low-$\ell$ HFI polarization power spectrum \texttt{EE}, and the lensing power spectrum. 
For $s$-wave annihilation with DM masses between MeV and GeV, the CMB angular power spectrum constraints are significantly stronger than those from DM indirect detection. 
In addition, Refs.~\cite{Slatyer:2009yq, Slatyer:2015jla} indicate that $f_c(z,\mathbf{x})$ in this redshift range can be approximated as a constant.
Consequently, for $s$-wave annihilation, an effective parameter constrained by CMB anisotropies is defined as $p_{\rm eff} \equiv f_c(z,\mathbf{x})\sv/m_\chi^2$, 
with the Planck 95\% upper limit of $p_{\rm eff} < 3.2 \times 10^{-28} \, \text{cm}^3 \, \text{s}^{-1} \, \text{GeV}^{-1}$ (Planck \texttt{TT+TE+EE+lowE+lensing+BAO})~\cite{Planck:2018vyg}. 
However, these simplifications are not accurate when considering a velocity-dependent annihilation cross-section. 
Therefore, based on the current Planck data and the future prospects of LiteBIRD~\cite{Suzuki2018TheLS} and CMB-S4~\cite{CMB-S4:2016ple,Abazajian:2019eic}, we derive comprehensive limits for $p$-wave and resonance annihilation scenarios in this work.

\subsection{Statistical framework}
\label{sec:statistics}

In this work, we utilize two types of likelihood functions in the analysis.
For blackbody spectral distortions, we adopt a step-function likelihood since experiments only provide upper limits.
Specifically, we use the 95\% upper limit of $|\mu| < 4.7 \times 10^{-5}$ from FIRAS, which means the likelihood drops to zero for any predicted $|\mu|$ exceeding this bound.\footnote{The 95\% upper limit is determined by a specific likelihood distribution. 
We use an effective step-function likelihood to ensure the signal strictly below the FIRAS 95\% confidence upper limit.}
Similarly, we apply upper limits to the likelihoods for all future experimental prospects, including the future measurements of blackbody spectral distortions and CMB anisotropies. 
For data measurements, such as the CMB angular power spectra (\texttt{TT}, \texttt{TE}, and \texttt{EE}), we use Gaussian likelihoods included in the numerical \texttt{MontePython}~\cite{Brinckmann2018MontePython3B}. To complete the likelihoods for CMB anisotropies, we incorporate the Planck 2018 data~\cite{Planck:2018vyg}, including high-$\ell$ (\texttt{TT}, \texttt{TE}, \texttt{EE}) spectra, low-$\ell$ temperature \texttt{TT} spectrum, low-$\ell$ polarization \texttt{EE} spectrum, lensing power spectrum, and BAO data~\cite{10.1093/mnras/stx721, Buen-Abad_2018, 10.1111/j.1365-2966.2011.19250.x, 10.1093/mnras/stv154}.

To present our results, we use two statistical methods: the ``marginal posterior'' (MP) method and the ``profile likelihood'' (PL) method. 
The MP approach integrates the probability density over nuisance parameters by marginalizing the posterior, commonly used in cosmology, especially within the $\Lambda$CDM framework. 
In contrast, the PL method is preferred for null signal searches due to parameter volume effects and prior dependencies in unconstrained likelihoods, and is widely applied in DM direct and indirect detection. 
For the MP method, we use the prior distributions as described in Ref.~\cite{Lewis:2013hha} on six cosmological parameters as nuisances---$\{\Omega_b h^2, \Omega_{\text{CDM}} h^2, 100\theta_s, \log(10^{10}A_s), n_s, \tau_{\text{reio}}\}$---informed by Planck precise central values and uncertainties.
However, we assign log-uniform priors for $\sv$, $b$, and $B_i B_f$ for three scenarios, respectively. 
For the PL method, we perform several fine scans in addition to the original Bayesian scan to improve coverage. 
For both methods, we exclude parameter regions where the accumulated probability exceeds 95\% for each fixed $m_\chi$.

\section{Numerical results}
\label{sec:Result}

\begin{table}[ht]
\begin{center}
\begin{tabular}{|c|c|c|c|c|}
\hline
 & $s$-wave & $p$-wave & resonance  \\
\hline\hline
$\mu$-distortion ($5 \times 10^4 \lsim z \lsim 2\times 10^{6}$) &  & $\checkmark$ & $3\times 10^4 \lsim z_\text{peak}\lsim 5\times10^{5}$ \\
\hline
$y$-distortion ($1000 \lsim z \lsim 5 \times 10^4$) &  &  & $3900 \lsim z_\text{peak}\lsim 2\times10^{4}$ \\
\hline
CMB anisotropies ($600 \lsim z \lsim 1000$) & $\checkmark$ &  & $600 \lsim z_\text{peak}\lsim 3900$ \\
\hline\hline
\end{tabular}
\caption{
Summary of the constraints from $\mu$-distortion, $y$-distortion, and CMB anisotropies on three types of annihilation cases. 
A checkmark ($\checkmark$) denotes the strongest constraint from the corresponding probe. 
For the resonance annihilation case (4\textsuperscript{th} column), specified $z_{\text{peak}}$ ranges are given, 
as the probes are sensitive to corresponding redshift ranges.
} 
\label{tab:constraints}
\end{center}
\end{table}
We compare the 95\% upper limits on $\mu$-distortion and $y$-distortion from FIRAS with those from CMB anisotropies in Planck.
We find that $\mu$-distortion provides a more stringent constraint for $p$-wave annihilation, 
due to its larger cross-section at higher redshifts ($5 \times 10^4 \lesssim z \lesssim 2 \times 10^6$).
For the resonance scenario,
we observe that when setting $z_\text{peak}$ above $5\times 10^5$ (for 1 GeV DM; this threshold increases for lower masses), the cross-section becomes sufficiently suppressed to evade current $\mu$-distortion limits from FIRAS observations.
This suppression arises because increasing $z_\text{peak}$ corresponds to a larger $\xi$ (i.e., a smaller cross-section) for fixed values of $m_\chi$ and $T_\text{kd}$, as derived in Eq.~\ref{equ:z peak}.
As we descend resonance peak redshifts $3900 \lsim z_\text{peak}\lsim 2\times10^{4}$, $y$-distortion constraints gain dominance due to enhanced energy injection during the critical $1000\lsim z \lsim 5\times 10^4$ window.
Finally, we lower $z_\text{peak}\lsim 3900$, Planck CMB anisotropy measurements supersede spectral distortion constraints as energy deposition increasingly overlaps with the recombination epoch $600\lsim z \lsim 1000$.
The most stringent limits for the three annihilation scenarios are summarized in Table.~\ref{tab:constraints}.

\begin{figure*}[ht]
\centering
\includegraphics[width=8cm]{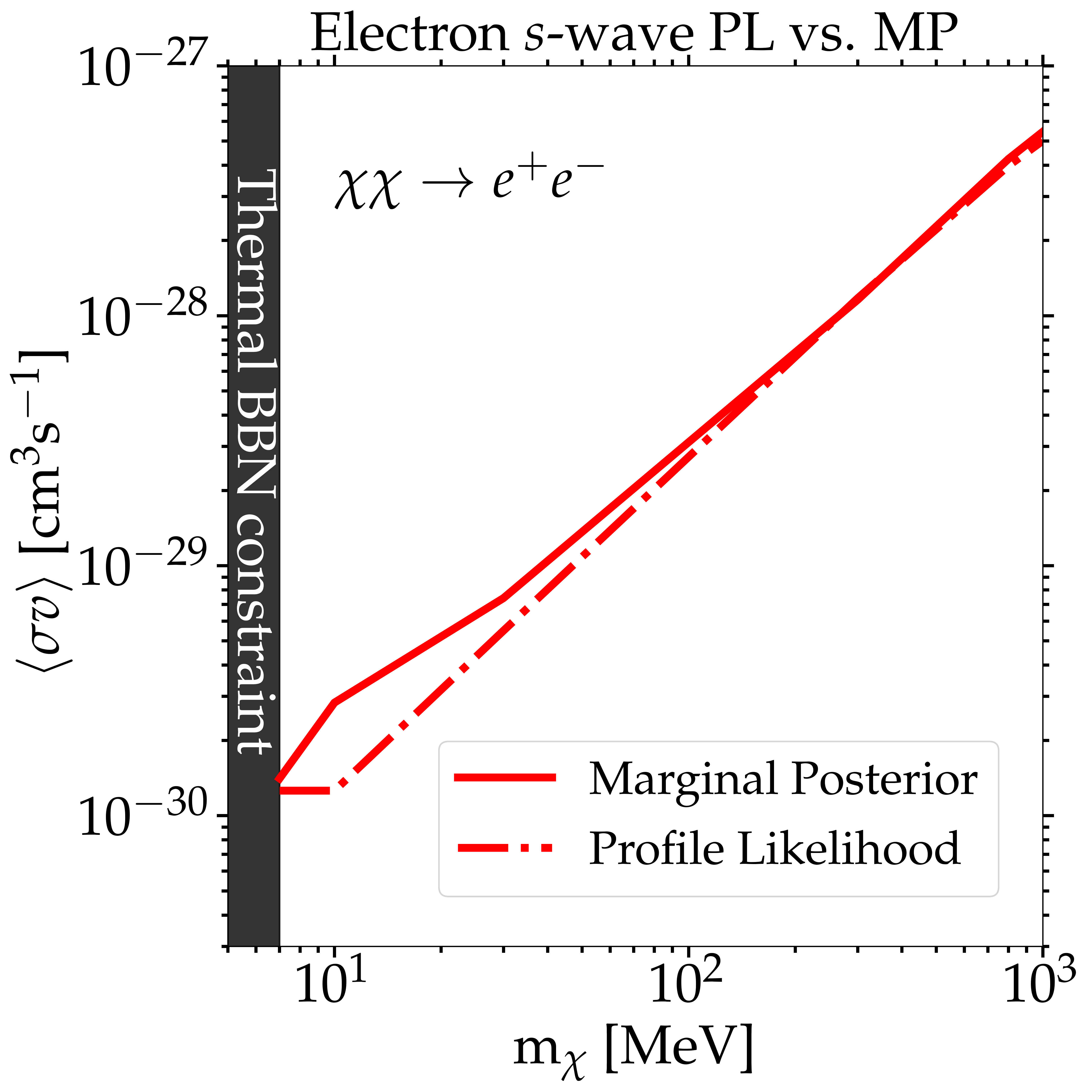}
\includegraphics[width=8cm]{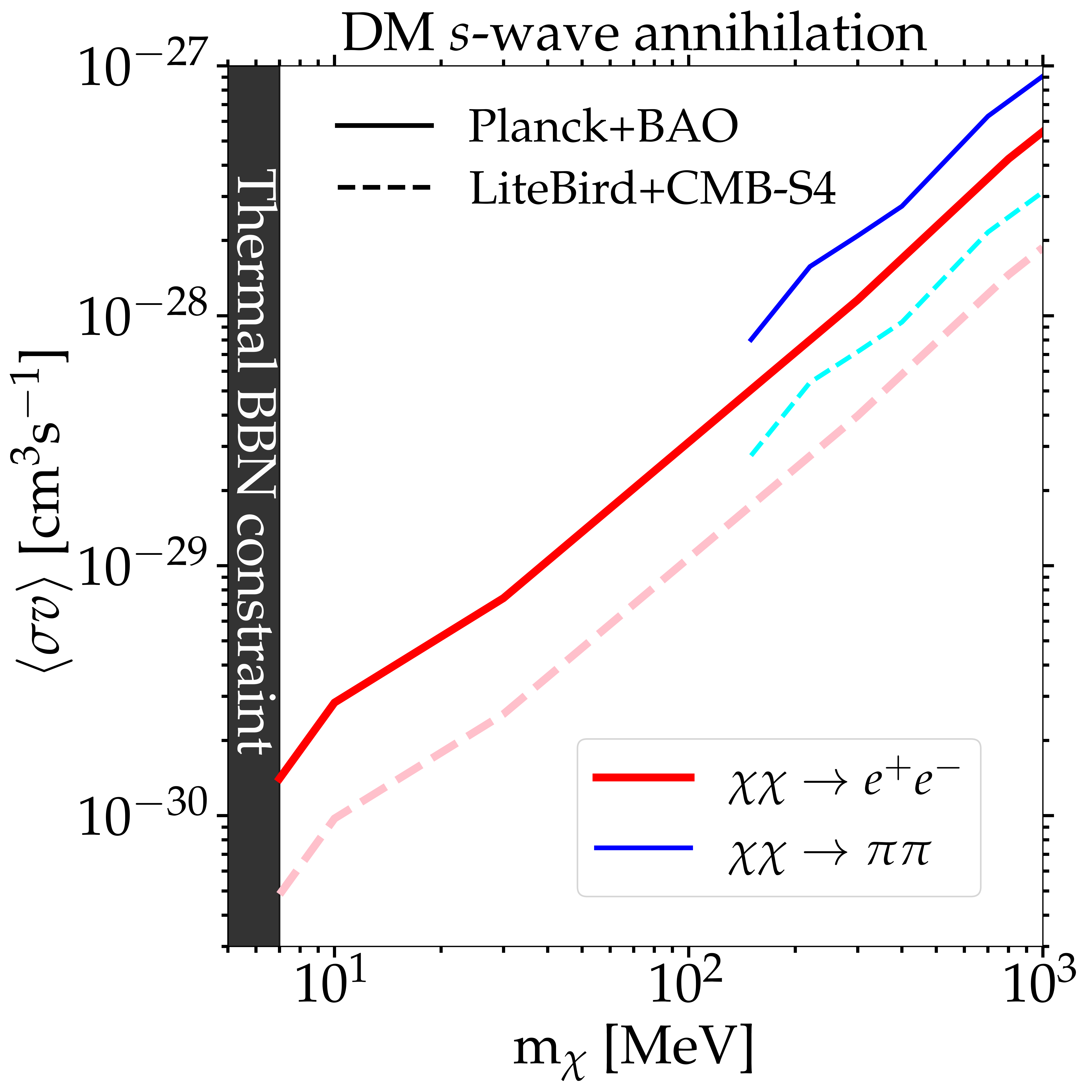}
 \caption{Comparison of the 95\% upper limits on $\sv$ for $s$-wave annihilation based on the \texttt{Planck+BAO} likelihood. 
 The left panel shows limits from the MP (red solid line) and PL (red dash-dotted line) methods for $e^+e^-$ final state. 
 The right panel compares the $e^+e^-$ (red) and $\pi\pi$ (blue) final states for \texttt{Planck+BAO} (solid lines) and future LiteBIRD and CMB-S4  experiments~\cite{Fu:2020wkq}.}
 \label{fig:swave}
\end{figure*}

Fig.~\ref{fig:swave} displays the 95\% upper limits on the ($m_\chi$, $\sv$) plane for the $s$-wave annihilation scenario, derived from the \texttt{Planck+BAO} likelihood. 
In the left panel, the limits for the $e^+e^-$ final state are compared between PL (red dash-dotted line) and MP (red solid line) method. 
The right panel compares the $e^+e^-$ (red) and $\pi\pi$ (blue) final states for the \texttt{Planck+BAO} likelihood (dark color) and the future LiteBIRD and CMB-S4 prospects~\cite{Fu:2020wkq} (light color). 
The black regions indicate the BBN exclusion of thermal DM particles with masses below $7\mev$~\cite{Depta_2019,Sabti_2021}.  
Hence, we set $7\mev$ as the lower mass limit for the electron channel.

The left panel of Fig.~\ref{fig:swave} shows that the PL result is stronger than the MP result, as expected. 
The MP method averages over all parameters, smoothing out nuisance effects and leading to less stringent constraints. 
In contrast, the PL method maximizes the likelihood for each parameter, providing sharper and more stringent limits on $\sv$. 
In the right panel of Fig.~\ref{fig:swave}, a lower mass limit of 150 MeV is set for the $\pi\pi$ final state, 
based on the rest mass of the $\pi^{\pm}$ (139.57 MeV) and $\pi^0$ (134.98 MeV). 
Our $e^+ e^-$ limits agree with previous studies~\cite{PhysRevD.102.103005}, while future experiments improve the constraints by a factor of approximately 3.
Note that blackbody spectral distortion limits are not as powerful as the \texttt{Planck+BAO} likelihood for probing the $s$-wave annihilation, 
even with the most advanced future blackbody spectral distortions missions~\cite{10.1093/mnras/stt1733,Fu:2020wkq}.


\begin{figure*}[ht]
\centering
\includegraphics[width=8cm]{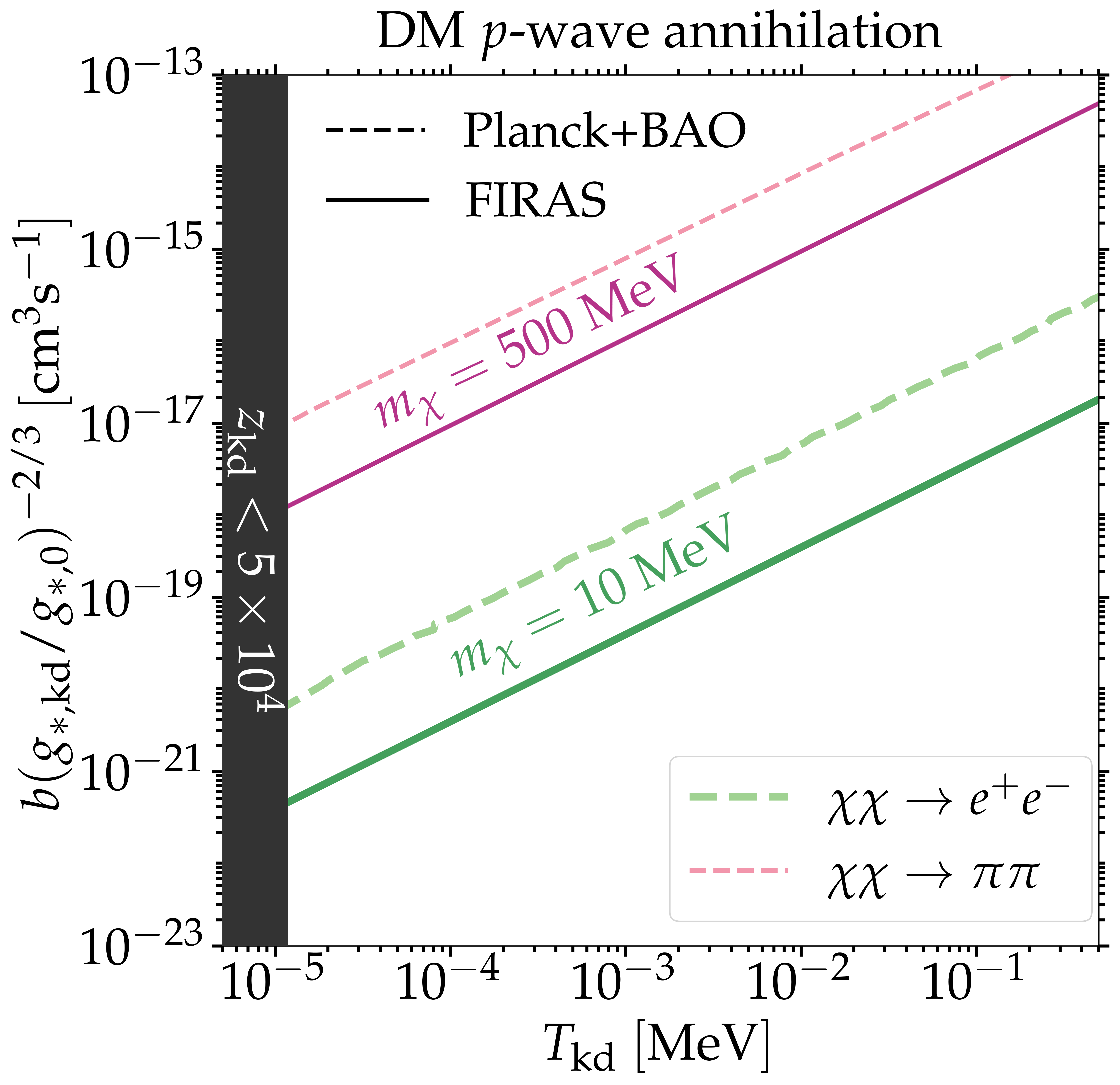}
\includegraphics[width=8cm]{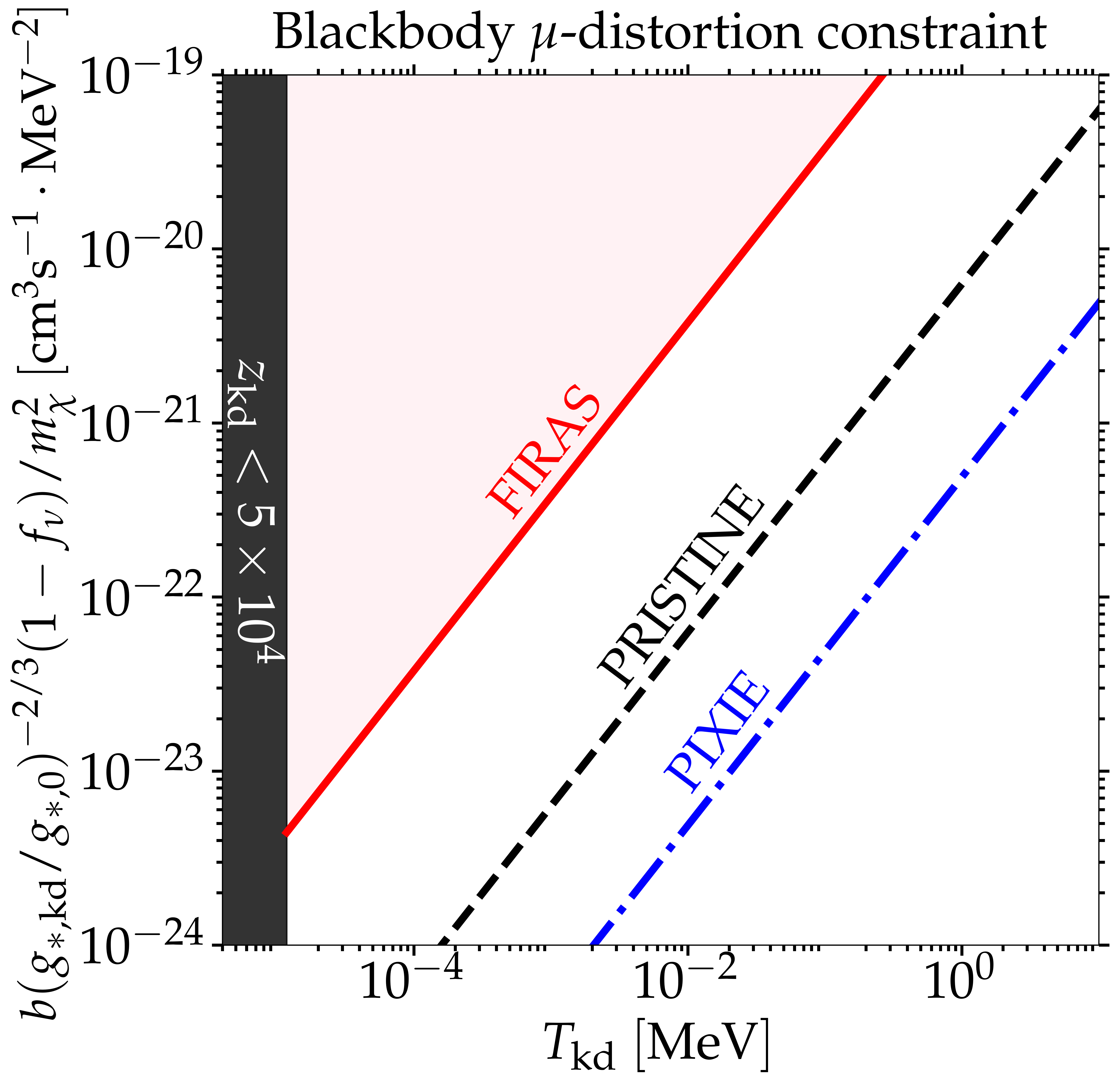}
\caption{The upper limits projected on the $T_{\rm kd}$ versus $b(g_{*,\text{kd}}/g_{*,0})^{-2/3}$ plane (left) and the $T_{\rm kd}$ versus $b(g_{*,\text{kd}}/g_{*,0})^{-2/3}(1-f_\nu)/m_\chi^2$ plane (right).
As shown in Tab.~\ref{tab:constraints}, the $\mu$-distortion data cannot probe the black region, 
where $z_{\rm kd} < 5 \times 10^4$, corresponding to $T_{\rm kd} < 1.18 \times 10^{-5}\ \mathrm{MeV}$.
In the left panel, the solid lines show the 95\% upper limit of FIRAS $\mu$-distortion (stronger than $y$-distortion) constraints for $m_\chi=10\mev$ (green line) and $m_\chi=500\mev$ (magenta line). 
Their corresponding dashed lines represent the 95\% upper limits of $b(g_{*,\text{kd}}/g_{*,0})^{-2/3}$ derived from the \texttt{Planck+BAO} likelihood, by using the MP method.
In the right panel, the red solid, black dashed, and blue dash-dotted lines represent the FIRAS, PRISTINE, and PIXIE the 95\% upper limit of $\mu$-distortion constraints, respectively.}
\label{fig:pwave}
\end{figure*}

Fig.~\ref{fig:pwave} shows the constraints for $p$-wave DM annihilation.
The left panel displays the relation between $T_{\rm kd}$ and $b(g_{*,\text{kd}}/g_{*,0})^{-2/3}$.
The black region, where $T_{\rm kd} < 1.18 \times 10^{-5}~\mathrm{MeV}$ (corresponding to $z_{\rm kd} < 5 \times 10^4$), 
represents the $\mu$-distortion insensitivity region, also see Tab.~\ref{tab:constraints}.
In the left panel, the FIRAS $\mu$-distortion constraints (solid lines) are stronger than those from the \texttt{Planck+BAO} likelihood (dashed lines), as the $p$-wave cross-section increases with redshift, making the $\mu$-distortion constraint stronger than the $y$-distortion one~\cite{Li2024ObservabilityOC}. 
Similarly, BBN and light-element abundances can yield stronger limits than the FIRAS constraints.
Although the PL result derived by the \texttt{Planck+BAO} likelihood is stronger than the MP result, it remains weaker than the FIRAS constraints.
In addition to the homogeneous DM background contribution, structure formation at $z<50$ enhances the squared number density and velocity dispersion of DM, which could significantly improve constraints on $p$-wave annihilation \cite{Diamanti2013ConstrainingDM}.
For 1 GeV DM annihilating into electrons with $T_\text{kd}=2.02\mev$, the result from~\cite{Diamanti2013ConstrainingDM} derives a 95\% upper limit of $b=1.8\times 10^{-16}\, \text{cm}^3 \, \text{s}^{-1}$, depending on halo model assumption.
In comparison, our model-independent constraints from FIRAS and PIXIE yield less stringent limits of $b < 1.64\times 10^{-12}\, \text{cm}^3 \, \text{s}^{-1}$ and $b < 2.14\times 10^{-15}\, \text{cm}^3 \, \text{s}^{-1}$, respectively, without relying on specific DM clustering assumption.

The right panel of Fig.~\ref{fig:pwave} illustrates the relation between $T_{\rm kd}$ and $b(g_{*,\text{kd}}/g_{*,0})^{-2/3}(1-f_\nu)/m_\chi^2$,  derived from the $\mu$-distortion constraints for DM $p$-wave annihilation.
Here, $f_\nu$ is the fraction of DM annihilation energy that goes into neutrinos. 
For DM annihilation into $e^+e^-$, $f_\nu$ is zero, while for charged pions, $f_\nu$ is approximately $0.47$.
Due to the linearity of upper limits, we can parameterize the 95\% constraints from FIRAS, PRISTINE, and PIXIE as
\begin{equation}\label{equ:pwave formula}
    \frac{T_\text{kd}}{1~\text{MeV}} \ge h_i \times b
    \times \left(\frac{g_{*,\text{kd}}}{g_{*,0}}\right)^{-2/3}
    \times (1-f_\nu) \times \left(\frac{1~\text{MeV}}{m_\chi}\right)^2,
\end{equation}
where $i=\{\text{FIRAS,~PRISTINE,~PIXIE}\}$ and $h_{\text{FIRAS}} = 2.643 \times 10^{18}$, $h_{\text{PRISTINE}} = 1.589 \times 10^{20}$, and $h_{\text{PIXIE}} = 2.045 \times 10^{21}$. This \textit{model-independent} inequality, Eq.~\eqref{equ:pwave formula}, is useful for testing arbitrary $p$-wave DM models.
As calculations using this formula and comparison with BBN limits~\cite{Braat:2024khe}, 
the future PIXIE mission's constraint on $p$-wave is comparable to BBN, while Super-PIXIE will surpass the BBN limits.


\begin{figure*}[htp]
\centering
\includegraphics[width=8cm]{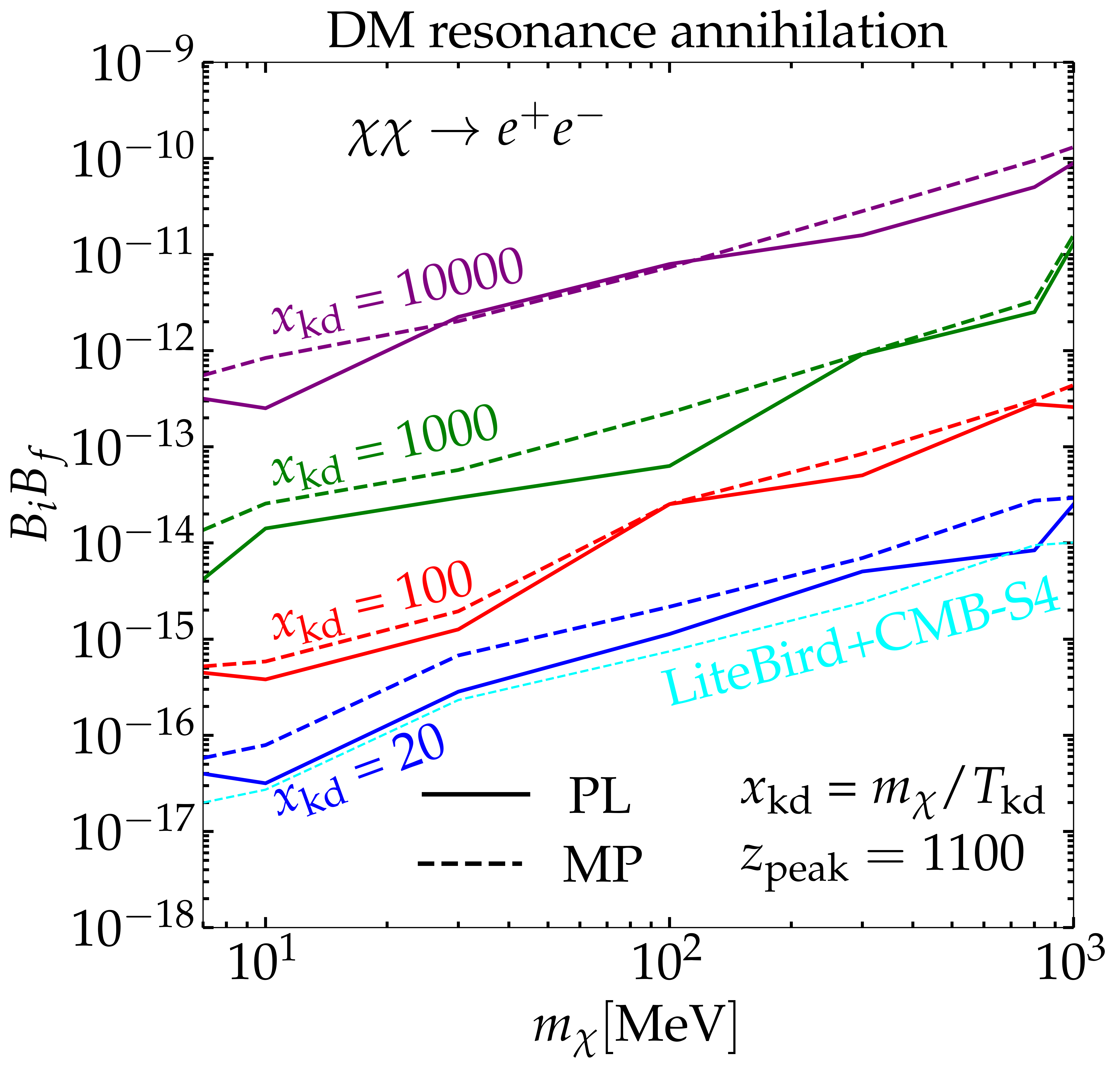}
\includegraphics[width=8.1cm]{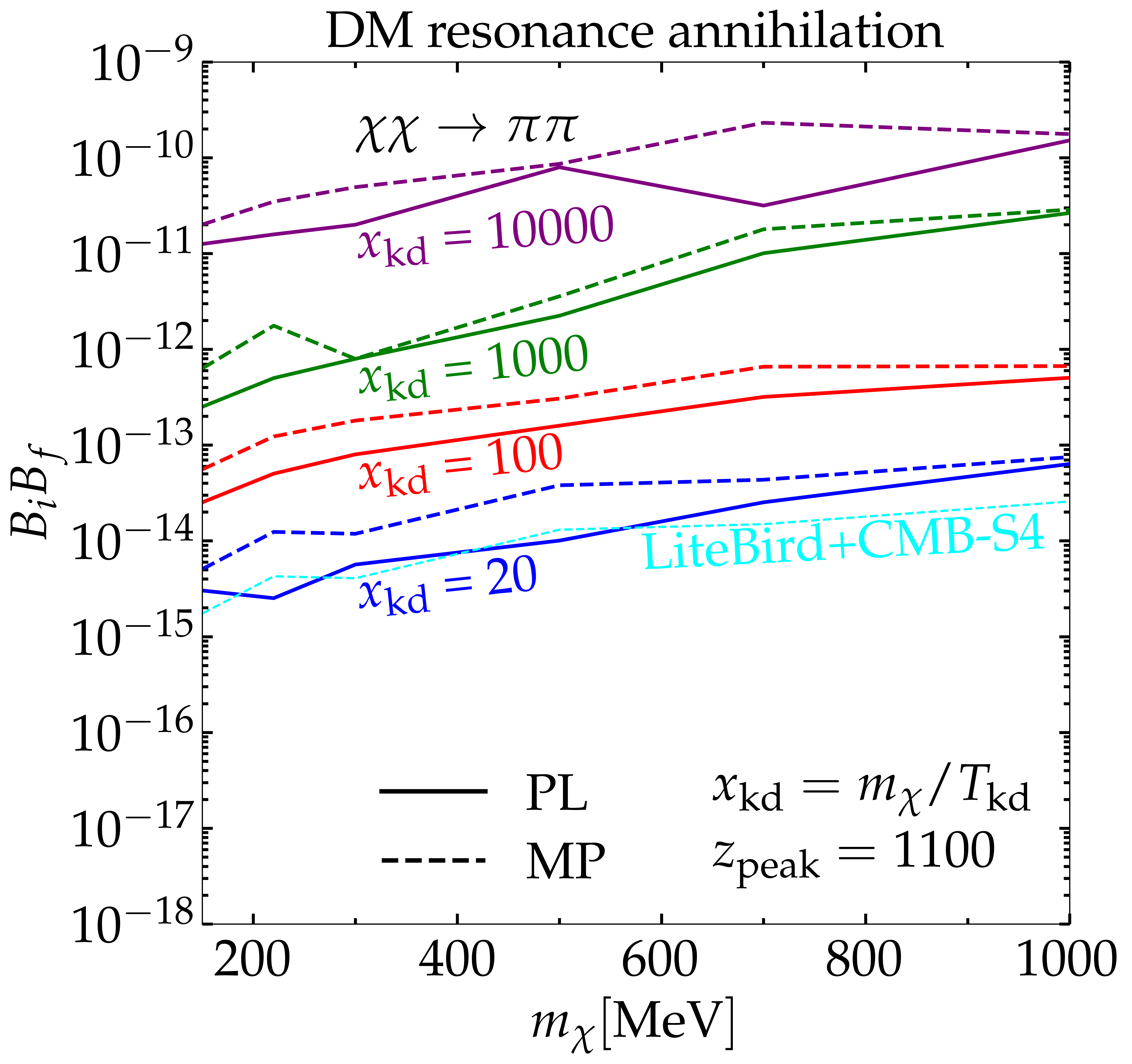}
\caption{The $95\%$ upper limits of the DM resonance annihilation decay branching ratio product, based on the \texttt{Planck+BAO} likelihood. 
The left panel shows the electron channel, while the right panel shows the pion channel. The solid lines use the PL method, and the dashed lines use the MP method. The values of $x_{\text{kd}} \equiv m_\chi/T_{\text{kd}} $ are $\{20, 100, 1000, 10000\}$, from bottom to top. 
The thin cyan dashed lines are derived using the configurations $x_{\text{kd}} = 20$, \texttt{LiteBIRD+CMB-S4}~\cite{Fu:2020wkq}, and the MP method.
} 
\label{fig:resonance}
\end{figure*}

Fig.~\ref{fig:resonance} shows the 95\% upper limits on $B_i B_f$ for the resonance annihilation scenario, derived from the \texttt{Planck+BAO} likelihood. 
We choose the resonance peak at $z_{\rm peak} = 1100$ as our benchmark.
The limits, obtained using the PL method (solid lines) and the MP method (dashed lines),
are shown for the $e^+ e^-$ (left panel) and $\pi\pi$ (right panel) final states.
The values of $x_{\text{kd}} \equiv m_\chi / T_{\text{kd}}$ are set to 20 (blue lines), $10^2$ (red lines), $10^3$ (green lines), and $10^4$ (purple lines).
The thin cyan dashed lines are derived by future LiteBIRD and CMB-S4 sensitivities~\cite{Fu:2020wkq} using the configurations $x_{\text{kd}} = 20$ and the MP method.
Note that smaller values of $x_{\text{kd}}$ correspond to earlier kinetic decoupling (larger $T_{\text{kd}}$) for the same DM mass. 
As shown in Eq.~\eqref{equ:z peak}, larger $T_{\text{kd}}$ results in smaller $\xi$, which strengthens the constraints on $B_i B_f$. 
For reference, we compute $\sv$ at $1+z = z_\text{peak} = 1100$ for a given $B_i B_f$. Taking $m_\chi = 1 \, \text{GeV}$ and $x_\text{kd} = 20$ (blue dashed line), the upper limits
$B_i B_f = 2.94 \times 10^{-14}$ (e$^+ e^-$) and $B_i B_f = 7.51\times 10^{-14}$ ($\mu^+ \mu^-$)
correspond to annihilation cross-sections $\sv$ of $2.24 \times 10^{-27} \, \text{cm}^3 \, \text{s}^{-1}$ and $5.63 \times 10^{-27} \, \text{cm}^3 \, \text{s}^{-1}$, respectively, at $1+z = z_\text{peak} = 1100$.

We comment on our results for $B_i B_f$ using the example of 100 MeV DM annihilation through dark Higgs into $e^+e^-$~\cite{Chen:2024njd}. 
For $100\mev$ DM annihilating into $e^+e^-$ shown in Fig.~\ref{fig:resonance}, 
we obtain $B_f < 10^{-16}$ under reasonable assumptions ($B_i \approx 1$ and $x_\text{kd} = 20$). 
The constraints on $B_f$, which scale as $B_f \propto \sin^2\theta$~\cite{Chen:2024njd}, imply $\sin\theta < 10^{-14}$ with $\xi = 2.02 \times 10^{-16}$. 
These limits are significantly stronger than those from the DarkSide-50 DM direct detection experiment~\cite{DarkSide-50:2023fcw}, as shown in~\cite{Chen:2024njd}. 
Note that our result in Fig.~\ref{fig:resonance} assumes $z_{\text{peak}} = 1100$.

\begin{figure*}[htb]
\centering
\includegraphics[width=8cm]{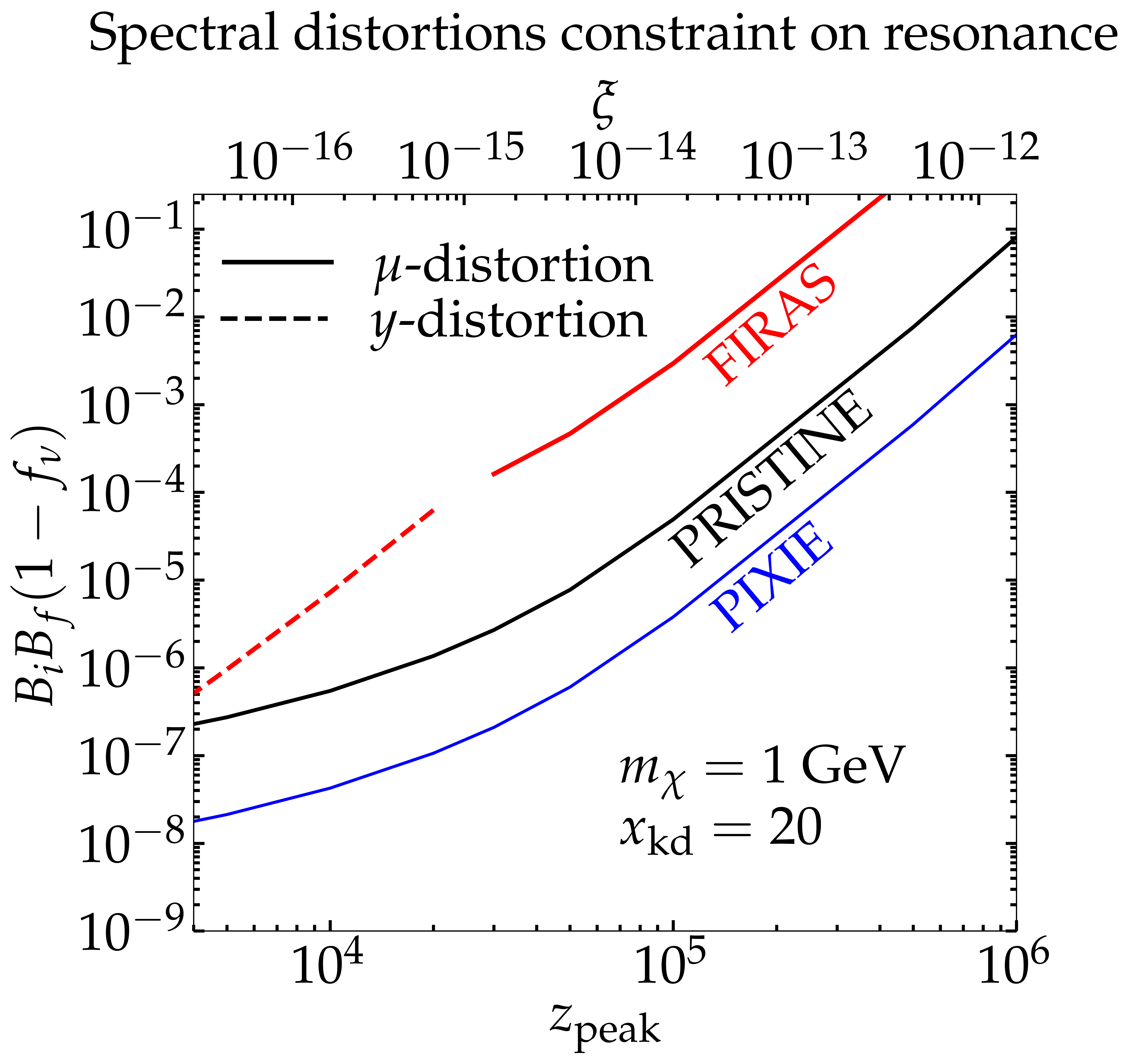}
\includegraphics[width=8cm]{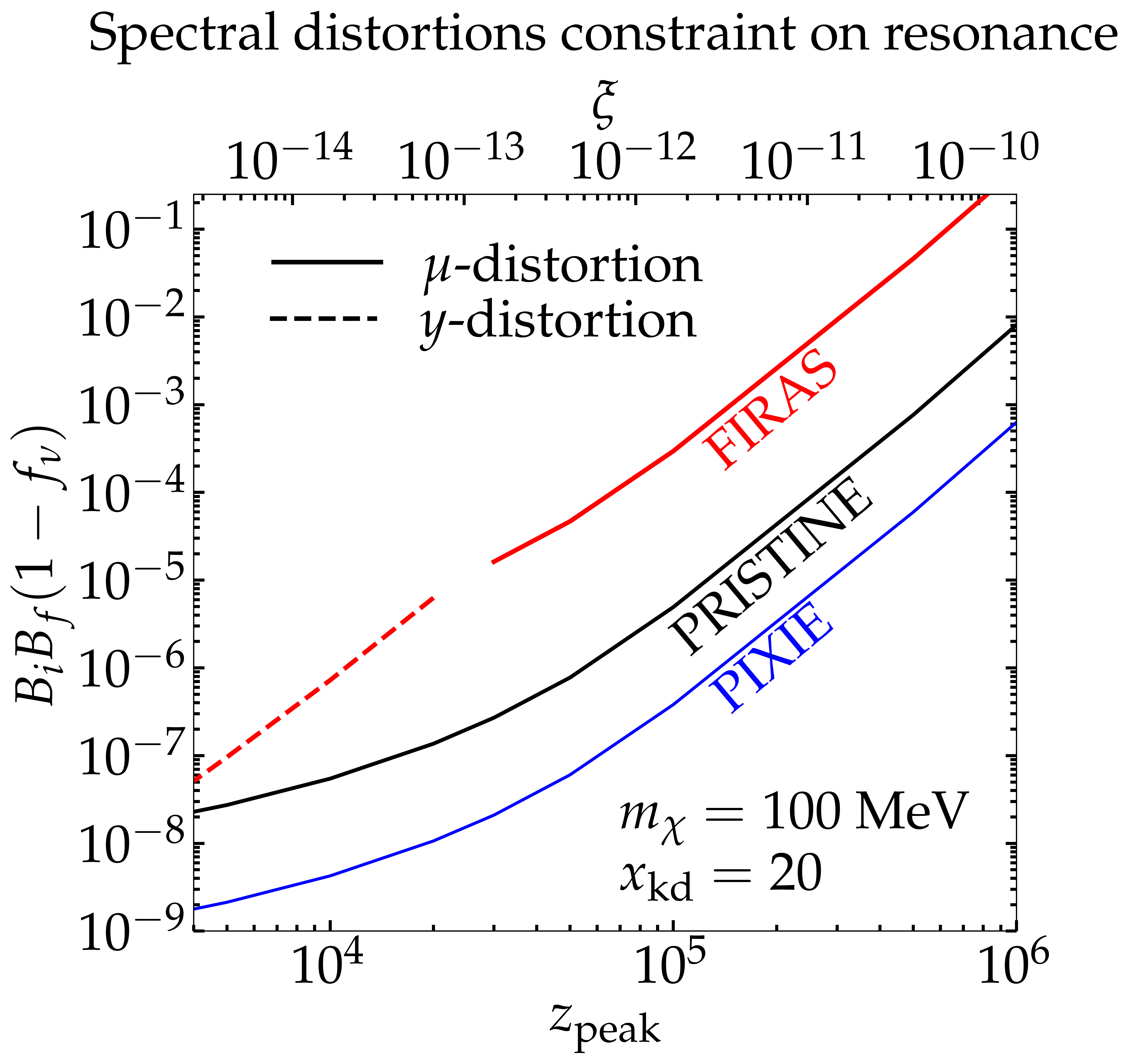}
\caption{The 95\% upper limits on the DM resonance annihilation branching ratio product $B_i B_f$, based on the spectral distortions likelihood, are shown as a function of $z_{\text{peak}}$ (bottom axis) and corresponding $\xi$ (top axis). The solid and dashed  lines indicate the sensitivity limits from $\mu$-distortion and $y$-distortions, respectively. Current constraints from FIRAS (red) and projected sensitivities from PRISTINE (black) and PIXIE (blue) are shown. The benchmark parameters used in the left and right plots are $m_\chi = 1\,\mathrm{GeV}$ and $m_\chi = 100\,\mathrm{MeV}$,  respectively, with $x_{\text{kd}} = 20$.}
\label{fig:resonance_SD}
\end{figure*}

Fig.~\ref{fig:resonance_SD} shows the constraints on the resonance branching ratios $B_i B_f$ from spectral distortion measurements, as a function of the resonance peak redshift $z_{\text{peak}}$ (bottom axis) and the corresponding coupling parameter $\xi$ (top axis).
The left panel assumes $m_\chi = 1 \, \mathrm{GeV}$, while the right panel assumes $m_\chi = 100 \, \mathrm{MeV}$, both with $x_{\text{kd}} = 20$.
Solid lines represent $\mu$-distortion constraints, while dashed lines represent $y$-distortion constraints. The red lines indicate current FIRAS limits, while black and blue lines show projected sensitivities for PRISTINE and PIXIE, respectively.  
For high peak redshifts $z_{\text{peak}} \gtrsim 5 \times 10^5$ (for 1 GeV DM), the cross-section suppression from large $\xi$ makes $\mu$-distortion constraints from FIRAS ineffective.
This threshold shifts to higher $z_{\text{peak}}$ for lighter DM, as seen in the 100 MeV case (right panel).
However, as $z_{\text{peak}}$ decreases below $5 \times 10^5$, $\mu$-distortion limits tighten, with PIXIE improving sensitivity by three orders of magnitude over FIRAS.
Once $z_{\text{peak}}$ falls below $2 \times 10^4$, $y$-distortion constraints (dashed lines) dominate.~\footnote{We consider only $y$-distortions from $z > 1000$, excluding contributions from the reionization epoch. 
This results in a more conservative upper limit on the DM annihilation cross-section, 
as our $y$-distortion estimate is lower than the full redshift integration.}

For even lower redshifts $z_{\text{peak}} \lesssim 3900$, CMB anisotropies constraints (e.g., at $z_{\text{peak}}=1100$ as shown in Fig.~\ref{fig:resonance}) surpass spectral distortions, as discussed in the main text.
The right panel shows that for 100 MeV DM, $\xi$ is larger than for 1 GeV DM at the same $z_{\text{peak}}$, which would nominally weaken the constraints. However, the higher number density of lighter DM ($n_\chi \propto m_\chi^{-1}$) compensates for this effect, leading to tighter bounds.
These results highlight the complementary roles of spectral distortions (between high-to-intermediate redshifts) and CMB anisotropies (at lower redshifts) in probing resonant DM annihilation.

\section{Conclusion}
\label{sec:summary}

In this work, we have systematically investigated the constraints imposed by the Planck, BAO, and FIRAS likelihood on sub-GeV DM  annihilation, considering both $e^{+}e^{-}$ and $\pi\pi$ final states through $s$-wave, $p$-wave, and resonance processes. 
The comparison of constraint strength between FIRAS and Planck CMB anisotropies is summarized in Table~\ref{tab:constraints}.

For the $s$-wave annihilation scenario, we have provided constraints for both the $e^{+}e^{-}$ and $\pi\pi$ channels. 
We find that the PL result is stronger than the MP result, as expected. 
We have further extended the analysis to include the pion channel and future experiments, such as LiteBIRD and CMB-S4, both derived using the MP method. 
Our results for the $e^{+}e^{-}$ channel agree with previous studies, while the inclusion of the $\pi\pi$ channel provides new insights into the constraints on DM annihilation with $m_\chi\sim \mathcal{O}(100)\mev$.

In the $p$-wave annihilation scenario, we derived a model-independent inequality (Eq.~\eqref{equ:pwave formula}) that parameterizes the $95\%$ upper limits from FIRAS, PRISTINE, and PIXIE.
Under the on-the-spot approximation during the $\mu$-distortion redshift range ($5 \times 10^4 \lesssim z \lesssim 2 \times 10^6$), the $\mu$-distortion constraints depend not only on the DM mass $m_\chi$ but also on the annihilation fraction into neutrinos $f_\nu$. This is because only the non-neutrino portion of the DM annihilation energy is deposited into the plasma, heating the photons and generating spectral distortions.

For $p$-wave annihilating DM with masses between 10 MeV and 1 GeV, the thermal relic density requires $b \sim 4 \times 10^{-25}~\text{cm}^3/\text{s}$ to $6 \times 10^{-25}~\text{cm}^3/\text{s}$~\cite{Diamanti2013ConstrainingDM}, as calculated numerically~\cite{Duan:2024urq}.
Considering a $p$-wave thermal relic DM benchmark with parameters $m_\chi = 10~\text{MeV}$, $b = 4 \times 10^{-25}~\text{cm}^3/\text{s}$, and $f_\nu = 0$, the kinetic decoupling occurs at $z_{\text{kd}} = 5 \times 10^{4}$. 
This value of $z_{\text{kd}}$ results in a maximum CMB $\mu$-distortion amplitude of $6.15 \times 10^{-8}$, which is below current FIRAS sensitivity limit ($|\mu| < 4.7 \times 10^{-5}$). However, future experiments like PIXIE sensitivity limit ($|\mu| < 8 \times 10^{-8}$ at 95\% confidence level) and Super-PIXIE (detection thresholds $\mu = 2\times 10^{-8}$ at 3$\sigma$) are sensitive to this predicted distortion level. Particularly, Super-PIXIE is expected to surpass current BBN limits, providing significantly more stringent constraints on $p$-wave DM annihilation.

For the resonance annihilation scenario, the strength of the constraints depends critically on the $z_\text{peak}$.  At high redshifts $z_\text{peak}\gtrsim 3900$, FIRAS spectral distortions provide the most stringent limits, and
spectral distortion constraints demand a fine-tuned parameter $\xi$.
Considering the model of an SM singlet Majorana DM interacting with the SM sector via an SM singlet real scalar boson, as discussed in Ref.~\cite{Chen:2024njd}, the viable parameter space for a DM mass of 100 MeV yields the correct relic density for $\xi \sim 3 \times 10^{-9}$. Meanwhile, $B_i B_f$ remains unconstrained between $5 \times 10^{-7}$ and $5 \times 10^{-4}$. 
While at lower redshifts $z_\text{peak}\lesssim 3900$, Planck CMB anisotropies dominate.
In this work, we focused on the case where the resonance peak occurs at $z_{\rm peak}=1100$.
Under this assumption, we calculated the constraints on the coupling coefficient $B_{i}B_{f}$ for both $e^{+}e^{-}$ and $\pi\pi$ final states. 
Our results indicate that the constraints on $\sin\theta$ are significantly more stringent than those from the  DarkSide-50 experiment, 
highlighting the power of CMB observations in probing DM properties.

In summary, our analysis demonstrates the complementary strengths of spectral distortions and CMB anisotropies constraints in probing sub-GeV DM annihilation. 
The inclusion of the $\pi\pi$ channel and the use of the PL method provides new and more stringent limits on $s$-wave annihilation. 
The model-independent inequality for $p$-wave annihilation and the future prospects of Super-PIXIE offer promising avenues. 
Finally, the resonance scenario, even under the weakest assumptions, provides constraints that surpass those from direct detection experiments.

\section*{Acknowledgments}
We sincerely thank Chi Zhang for the \texttt{MontePython} discussion and Meiwen Yang for the resonance discussion. This work is supported by the National Key Research and Development Program of China (No. 2022YFF0503304), the Project for Young Scientists in Basic Research of the Chinese Academy of Sciences 
(No. YSBR-092), and the Jiangsu Province Post Doctoral Foundation (No. 2024ZB713).

\newpage
\appendix

\section{The root-mean-square velocity and $p$-wave annihilation}
\label{app:p-wave}

We can replace the parameter $b$ by a reference velocity $v_{\texttt{100}}$ and its corresponding cross-section $\svloo$. 
Here, the suffix $\texttt{100}$ indicates that the root-mean-square velocity satisfies $\langle v_\chi^2 \rangle = (100~{\rm km/s})^2\equiv v_{\texttt{100}}^2$, reflecting the DM dispersion velocity at the present.
Thus, the velocity-averaged $p$-wave annihilation cross-section can be expressed by scaling from $\svloo$, 
 \begin{equation}
 \label{eq:svpwave}
    \sv =\frac{\svloo}{v_{\texttt{100}}^2} \langle v_\chi^2 \rangle,~{\rm with}~b\equiv \frac{\svloo}{v_{\texttt{100}}^2} . 
\end{equation}

After DM kinetic decoupling and becoming non-relativistic, its temperature $T_\chi\propto (1+z)^2$. Using equipartition of energy for an ideal gas, $\langle v_\chi^2 \rangle=3T_\chi/m_\chi$, we have 
\begin{equation}\label{equ:T(1+z)}
   \frac{\langle v_\chi^2 \rangle}{v_{\texttt{100}}^2}=\frac{T_\chi(z)}{T_{\texttt{100}}}=\bigg(\frac{1+z}{1+z_{\texttt{100}}}\bigg)^2.
\end{equation}
where $z_{\texttt{100}}$ is the redshift when $\langle v_\chi^2 \rangle=v_{\texttt{100}}^2$ and $T_{\texttt{100}}$ is the DM temperature at this redshift. 
By using Eq.~\eqref{equ:T(1+z)} and equipartition of energy, we make the substitution 
\begin{equation}\label{equ:z100}
    1+z_{\texttt{100}}=\frac{v_{\texttt{100}}}{c}(1+z_{\text{kd}})\bigg(\frac{m_\chi}{3T_\text{kd}}\bigg)^{1/2},
\end{equation}
where $c$ is the speed of light, and the redshift of DM kinetic decoupling is $z_\text{kd}$. 
Using the fact that DM was in thermal equilibrium with the CMB at the time of kinetic decoupling and conservation of entropy $s\propto (1+z)^{3}$, we can express $z_{\text{kd}}$ in terms of $T_{\text{kd}}$ and the CMB temperature in today $T_\text{CMB,0}=2.35\times 10^{-10}~\text{MeV}$ as 
\begin{equation}\label{equ:zkd}
    1+z_\text{kd}=\left(\frac{g_{*,\text{kd}}}{g_{*,0}}\right)^{1/3}
    \frac{T_\text{kd}}{T_\text{CMB,0}} \simeq 4.26 \times 10^9 \left(\frac{g_{*,\text{kd}}}{g_{*,0}}\right)^{1/3} \left(\dfrac{T_\text{kd}}{\text{MeV}}\right),
\end{equation}
where $g_{*,\text{kd}}$ and $g_{*,0}$ are the effective numbers of relativistic degrees of freedom at the redshift of DM kinetic decoupling and at present $(z=0)$, respectively.

Finally, we obtain the root-mean-square velocity by using Eq.~\eqref{equ:z100} and Eq.~\eqref{equ:zkd},   
\begin{equation}\label{equ:vrms}
    v_{\text{rms}}^2=1.66\times10^{-19}(1+z)^2
    \left(\frac{g_{*,\text{kd}}}{g_{*,0}}\right)^{-2/3}
    \left(\frac{1~\text{MeV}}{m_\chi}\right)
    \left(\frac{1~\text{MeV}}{T_{\text{kd}}}\right).
\end{equation}


\section{The velocity-averaged cross-section near resonance}\label{app:resonance}
The general formula for the scattering cross-section via a resonance ($\chi\chi\rightarrow\phi\rightarrow f\bar f$) is
\begin{equation}\label{app-equ:resonance sigma}
    \sigma=\frac{16\pi}{m_\phi^2\Bar{\beta_i}\beta_i}\frac{\gamma^2}{(-\xi+v_{\text{rel}}^2/4)^2+\gamma^2}B_iB_f.
\end{equation}
Considering that DM particles are non-relativistic at resonance, 
we can adopt the Maxwell-Boltzmann velocity distribution to compute velocity-averaged annihilation cross-section 
\begin{equation}\label{app-equ:resonance cross-section}
    \sv=\frac{1}{(2\pi v^2_{\text{rms}}/3)^3}\int \text{d}\Vec{v}_1\int \text{d}\Vec{v}_2~e^{-3(v_1^2+v_2^2)/(2v^2_{\text{rms}})}\times \sigma |\Vec{v}_1-\vec{v}_2|,
\end{equation}
where the root-mean-square velocity are given in Eq.~\eqref{equ:vrms}.

In the non-relativistic limit, the center-of-mass energy is $E^2_{\text{cm}}\simeq 4m_\chi^2+m_\chi^2v_{\text{rel}}^2$, 
and the relative velocity is $v_\text{rel}\simeq 2\beta_i$. 
We replace the velocities $\vec{v}_1$ and $\vec{v}_2$ with 
\begin{equation}
    \Vec{v}_{\text{cm}}=\frac{\Vec{v}_1+\Vec{v}_2}{2},~\Vec{v}_{\text{rel}}=\Vec{v}_1-\Vec{v}_2,
\end{equation} 
and take $\text{d}\Vec{v}_{\text{cm}} \text{d}\Vec{v}_{\text{rel}}=\text{d}\Vec{v}_1 \text{d}\Vec{v}_2$. 
Then, we can simplify Eq.~\eqref{app-equ:resonance cross-section} by using new velocity variables $\Vec{v}_{\text{cm}}$ and $\Vec{v}_{\text{rel}}$, 
\begin{equation}
\begin{aligned}
    \sv=
    &\frac{16\pi^2}{(2\pi v^2_{\text{rms}}/3)^3}\frac{32\pi}{m_\phi^2\Bar{\beta_i}}B_iB_f 
     \int \text{d}\Vec{v}_{\text{cm}}v^2_{\text{cm}}e^{-3v^2_{\text{cm}}/v^2_{\text{rms}}} \\
    & \int\text{d}\Vec{v}_{\text{rel}}v^2_{\text{rel}}e^{-3v^2_{\text{rel}}/(4v^2_{\text{rms}})}
       \frac{\gamma^2}{(-\xi+v_{\text{rel}}^2/4)^2+\gamma^2}.
\end{aligned}
\end{equation}
We can obtain the integration of $\Vec{v}_{\text{cm}}$ 
\begin{equation}
    \int \text{d}\Vec{v}_{\text{cm}}
    v^2_{\text{cm}}\exp{\Big[-3v^2_{\text{cm}}/v^2_{\text{rms}} \Big]}
    =\frac{1}{12}\sqrt{\frac{\pi}{3}}v^3_{\text{rms}}, 
\end{equation}
and use narrow width approximation
\begin{equation}
    \lim\limits_{\gamma/\xi \to 0}\frac{1}{(-\xi+v_{\text{rel}}^2/4)^2+\gamma^2}=
    \frac{\pi}{\gamma}\delta(-\xi+v_{\text{rel}}^2/4),
\end{equation}
to perform the integration of $\Vec{v}_{\text{rel}}$ as 
\begin{equation}
    \int\text{d}\Vec{v}_{\text{rel}}v^2_{\text{rel}}e^{-3v^2_{\text{rel}}/(4v^2_{\text{rms}})}
   \frac{\gamma^2}{(-\xi+v_{\text{rel}}^2/4)^2+\gamma^2}
   \simeq 4\sqrt{\xi}\exp{[-3\xi/v^2_{\text{rms}}]}\gamma\pi.
\end{equation}
Therefore, the velocity averaged cross-section of resonance is
\begin{equation}
\sv=\frac{576}{\sqrt{3}}
    \frac{\pi^{3/2}}{m_\phi^2}
    \frac{\gamma}{v^3_{\text{rms}}}
    \exp{[-3\xi/v^2_{\text{rms}}}]
    B_i B_f.
\end{equation}


\section{\texttt{DarkHistory} and \texttt{CLASS} Parameter Settings} \label{app:code}

There are two approaches to incorporating exotic energy deposition from \texttt{DarkHistory} into \texttt{CLASS}: either passing the calculated deposition function $f_c(z)$ as inputs to the \texttt{ExoCLASS} branch~\cite{Stocker:2018avm}, or directly inserting precomputed tables of the injection terms $\dot T^{\text{inj}}_m$ and $\dot x^{\text{inj}}_\text{HII}$ defined in Eq.~\eqref{dTdx}. 

In this work, we adopt the latter methodology by calculating the evolution of $\dot T^{\text{inj}}_m$ and $\dot x^{\text{inj}}_\text{HII}$ with exotic energy injection, then interfacing these results through the \texttt{HyRec}~\cite{PhysRevD.83.043513} recombination module implemented in the \texttt{ExoCLASS} framework.

To enable this implementation, we introduce a set of new DM parameters that govern the energy injection process:
\begin{description}[leftmargin=3cm,labelsep=0.5cm]
  \item[\texttt{DM\_annihilation\_DH\_flag}] Toggles the \texttt{DarkHistory} treatment (enabled/disabled).
  \item[\texttt{DM\_annihilation\_DH\_mass}] Specifies the dark matter particle mass.
  \item[\texttt{DH\_cross\_section\_exponent}] Controls logarithmic sampling of cross-sections.
  \item[\texttt{DH\_Tkd\_denominator}] Defines the kinetic decoupling parameter $x = m_\chi / T_\mathrm{kd}$.
  \item[\texttt{DM\_annihilation\_DH\_channel}] Selects the annihilation final state (e.g., $e^+e^-$, $\pi\pi$).
  \item[\texttt{DM\_annihilation\_DH\_mechanisms}] Specifies the annihilation mechanism (e.g., $s$-wave, $p$-wave, resonance).
\end{description}

The parameter settings in \texttt{DarkHistory} (except for DM-related parameters) are:
\begin{align*}
\mathtt{start\_rs}       &= \mathtt{3000}, \\
\mathtt{end\_rs}         &= \mathtt{4}, \\
\mathtt{reion\_switch}   &= \mathtt{False}, \\
\mathtt{struct\_boost}   &= \mathtt{None}, \\
\mathtt{elec\_method}    &= \mathtt{\text{\textquoteleft} \mathtt{new} \text{\textquoteright}}, \\
\mathtt{nmax}            &= \mathtt{20}.
\end{align*}
The remaining parameters retain their default values in \texttt{DarkHistory}.

The parameter settings (except for DM-related and cosmological parameters) in \texttt{CLASS} are:
\begin{align*}
\mathtt{recombination}       &= \mathtt{HyRec}, \\
\mathtt{non\_linear}         &= \mathtt{halofit}, \\
\mathtt{N\_ur}               &= \mathtt{2.0328}, \\
\mathtt{N\_ncdm}             &= \mathtt{1}, \\
\mathtt{m\_ncdm}             &= \mathtt{0.06}, \\
\mathtt{T\_ncdm}             &= \mathtt{0.71611}, \\
\mathtt{distortions\_verbose} &= \mathtt{2}.
\end{align*}
These non-cold dark matter parameter settings follow the \texttt{MontePython} convention to ensure the total effective neutrino degrees of freedom $N_{\text{eff}} = 3.046$ (consistent with Big Bang Nucleosynthesis and CMB measurements) while satisfying the cosmological mass-density relation $m_{\nu}/\Omega_{\nu}h^2 = 93.14  \text{eV}$ for the massive neutrino component. The remaining parameters retain their default values in \texttt{CLASS}.


\bibliography{reference.bib}

\end{document}